\documentclass[prb,showpacs,superscriptaddress,titlepage,amsmath,amssymb,twocolumn,floatfix]{revtex4-1}

\usepackage{tikz}
\usepackage{amsmath}
\usepackage{amssymb}
\usepackage{graphicx}
\usepackage{xcolor}
\usepackage{amsthm}
\usepackage{subfigure}
\usepackage{color}
\usepackage{soul}
\usepackage{bbold}
\usepackage{bibunits}

\begin{document}
\preprint{APS/123-QED}

\title{Accurate calculation of excitonic signatures in the absorption spectrum of BiSBr using semiconductor Bloch equations}
\author{J.~M.~Booth}
\email{jamie.booth@rmit.edu.au}
\affiliation{ARC Centre of Excellence in Exciton Science, RMIT University, Melbourne, Australia}
\affiliation{Theoretical Chemical and Quantum Physics, RMIT University, Melbourne, Australia}
\author{M.~V.~Klymenko}
\email{mike.klymenko@rmit.edu.au}
\thanks{These two authors contributed equally to this work.}
\affiliation{ARC Centre of Excellence in Exciton Science, RMIT University, Melbourne, Australia}
\affiliation{Theoretical Chemical and Quantum Physics, RMIT University, Melbourne, Australia}
\author{J.~H.~Cole}
\affiliation{ARC Centre of Excellence in Exciton Science, RMIT University, Melbourne, Australia}
\affiliation{Theoretical Chemical and Quantum Physics, RMIT University, Melbourne, Australia}
\author{S.~P.~Russo}
\affiliation{ARC Centre of Excellence in Exciton Science, RMIT University, Melbourne, Australia}
\affiliation{Theoretical Chemical and Quantum Physics, RMIT University, Melbourne, Australia}
\date{\today}

\begin{abstract}

In order to realize the significant potential of optical materials such as metal halides, computational techniques which give accurate optical properties are needed, which can work hand-in-hand with experiments to generate high efficiency devices. In this work a computationally efficient technique based on semiconductor Bloch equations (SBEs) is developed and applied to the material BiSBr. This approach gives excellent agreement with the experimental optical gap, and also agrees closely with the excitonic stabilisation energy and the absorption spectrum computed using the far more computationally demanding \textit{ab initio} Bethe-Salpeter approach. The SBE method is a good candidate for theoretical spectroscopy on large- or low dimensional systems which are too computationally expensive for an \textit{ab initio} treatment.
\end{abstract}

\maketitle
\section{Introduction}
The optical properties of metal-halide materials have been of considerable interest recently due to their high efficiency in photovoltaic applications, long exciton diffusion lengths, and relative stability of these properties to structural defects\cite{Stranks2013,Yang2015,Brandt2015,Fu2019,Shamsi2019}. New generations of high efficiency, low cost devices using their optical properties are promising not just for photovoltaics, but also water-splitting and other photo-catalytic applications and sensors\cite{Fu2019,Shamsi2019}. In order to realize this potential we need computational tools available to facilitate the development of new, efficient forms.  

Optical properties of materials are extremely computationally expensive to calculate \textit{ab initio} in comparison to properties such as structural parameters or band structures which may be amenable to Density Functional Theory\cite{Burke2012} or the GW Approximation\cite{Aryasetiawan1998}. The reason for this is that optical spectra are dominated by neutral excitations, and precise calculations need to accurately reproduce not only the electron interactions which dress excitations (the quasiparticle description) but then accurately reproduce the interactions between the dressed electrons and holes\cite{Onida2002}. 

This is a formidable computational task, and while approaches such as Time-Dependent Density Functional Theory (TDDFT\cite{Casida2012}) can provide information on excited states, only the more formal Bethe-Salpeter Equation (BSE) approach performed on top of GW calculations includes electron-hole interactions in an \textit{ab initio} manner\cite{Onida2002,Reining2002,Cunningham2018}. However, the extreme resource requirements of BSE calculations have stymied widespread uptake. A good summary of these approaches and related issues can be found in the review of Onida, Reining and Rubio\cite{Onida2002}.

Alternatively, in the case of excitons with large Bohr radii (Wannier-Mott excitons), an efficient model for optical response can be built using an effective mass model and semiconductor Bloch equations (SBEs) \cite{Haug2004_BEs}. This approach, being time-dependent and non perturbative is also amenable to the calculation of time-dependent linear and non-linear optical properties such as transient-absorption spectroscopy and other pump-probe experiments\cite{Katsch2020,Hannes2020}.

Recently it was shown that SBEs can be applied to interlayer excitons in van-der-Waals bonded transition-metal dichalcogenide layers\cite{Meckbach2018}. In that study, the model for the optical properties which was derived went beyond the effective mass approximation, and requires free parameters that can be obtained from DFT calculations. In this work, we attempt a similar approach in order to determine if it can be an effective and efficient alternative to much more computationally expensive \textit{ab initio} methods.

We chose to focus on the metal chalcohalide BiSBr, due to its intriguing properties: an optical gap in the visible range of the electromagnetic spectrum\cite{Ran2018}, synthesis at low-temperature and high photo-currents\cite{Kunioku2016}.

\begin{figure}[!t]
    \centering
    \subfigure[]{\includegraphics[width=0.6\linewidth]{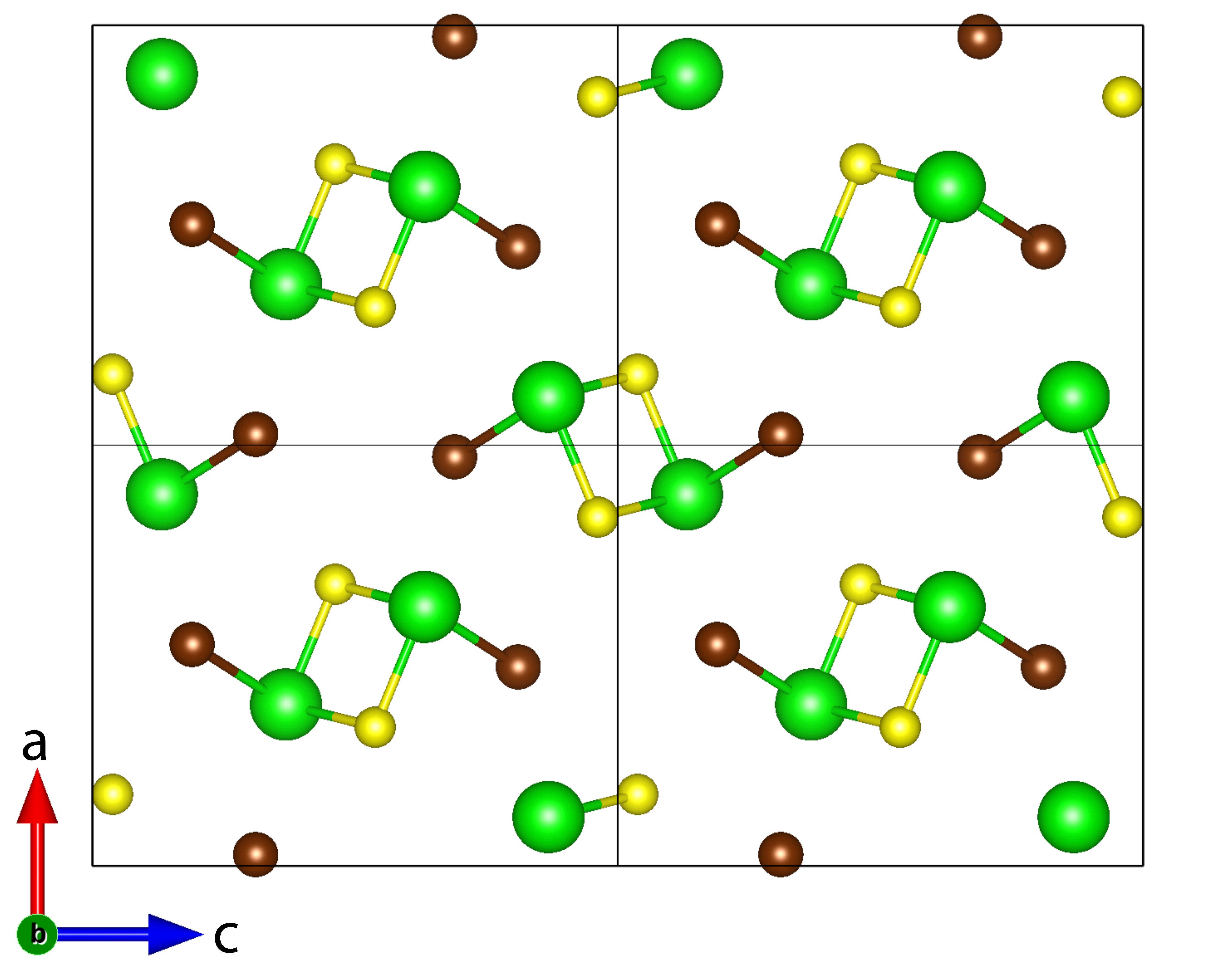}}
    \subfigure[]{\includegraphics[width=0.6\linewidth]{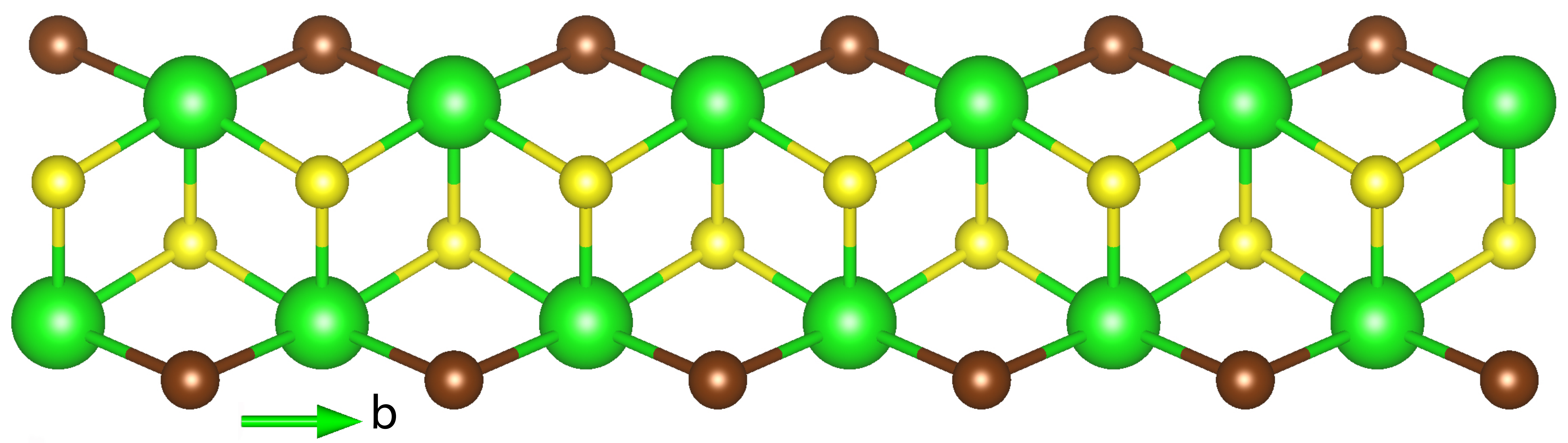}}
    \caption{(Color Online) a) View down the b-axis of the BiSBr Pnma crystal structure, showing the ``cluster" shape of the chains which make up the crystal. Bismuth atoms are green, bromine atoms are brown and sulfur atoms are yellow. b) Side view of one of the cluster chains of the previous perspective showing that the Bismuth atoms are arranged in an effectively one-dimensional structure.}
    \label{BiSBr}
\end{figure}

BiSBr crystallizes in a \textit{Pnma} structure\cite{Brandt2015}, two perspectives of which are presented in Figure \ref{BiSBr}. Figure \ref{BiSBr}a) shows a view down the crystal $\mathbf{b}$-axis which highlights the``cluster" nature of the BiSBr structure. In the centers of the two unit cells a Bi-S cluster is visible, which is weakly bound to the other parallel cluster by van der Waals-type bonding. These Bi-S units are linked down the $\mathbf{b}$-axis, as Figure \ref{BiSBr}b demonstrates. They form zig-zag chains of Bi atoms connected by sulphur and bromine atoms which are isolated from the other parallel chains. 

The quasi one-dimensional structure suggests a corresponding one-dimensional electronic structure, in which considerable confinement of excitons will occur, significantly enhancing their stabilities\cite{Rossi1999,Chen2005,Brus2010}. Optical spectroscopy measurements on single BiSBr crystals determined the neutral excitation gap to be 2.01 eV\cite{Jin1995}.

Despite the potential of metal halide materials for photovoltaic applications, rigorous \textit{ab initio} studies have been reported on only a handful of systems, such as TDDFT data\cite{Demchenko2016} and static hybrid DFT data\cite{Mosconi2016} on methyl-ammonium lead iodide (MA-PbI) and MA-PbX where X = I, Cl, Br respectively.

The purpose of this work is therefore to determine whether a far less computationally demanding approach based on semiconductor Bloch equations can, when used in concert with \textit{ab initio} electronic structure approaches such as the GW approximation, provide data which is in good agreement with full \textit{ab initio} methods, but far more amenable to large, or low-dimensional structures.

\section{Methods}
The optical properties of materials are contained in the macroscopic dielectric matrix $\epsilon_{M}(\omega)$, which  is obtained from the modified response function $\bar{\chi}$, via:
\begin{equation}
	\epsilon_{M}(\omega) = 1- \lim\limits_{\mathbf{q}\rightarrow 0}V_{\mathbf{G}=0}(\mathbf{q})\bar{\chi}_{\mathbf{G}=\mathbf{G}^{\prime}=0}(\mathbf{q};\omega)
\end{equation}
where $V_{\mathbf{G}=0}(\mathbf{q})$ is the Coulomb interaction, and $\mathbf{G},\mathbf{G}^{\prime}$ denote reciprocal lattice vectors.

Once the dielectric response is obtained the absorption spectrum can be calculated via:
\begin{equation}
	\alpha(\omega) = \frac{\omega\epsilon_{2}(\omega)}{\sqrt{\frac{1}{2}\big(\epsilon_{1}(\omega)+\sqrt{\epsilon_{1}^{2}(\omega)+\epsilon_{2}^{2}(\omega)}\big)}}
	\label{eq:Abs}
\end{equation}
where $\epsilon_{1}(\omega)$ and $\epsilon_{2}(\omega)$ are the real and imaginary parts of the frequency-dependent dielectric response respectively. 

In this work, three different approaches are used to calculate the optical response of BiSBr: Time-Dependent Density Functional Theory (TDDFT), solving the Bethe-Salpeter Equation, and an approach based on solving a system of semiconductor Bloch Equations (SBE). These three approaches differ significantly in their computational approach and we describe them in detail in the Supplementary Material.\cite{Booth_2021_SI} We give a brief description of the Seminconductor Bloch Equation calculations below.

\subsection{Semiconductor Bloch equations}

In this work we modify the semiconductor Bloch equations to take into account the local field effect in the directions perpendicular to the atomic chains. The derivation of the modified semiconductor Bloch equations (shown in detail in the Supplementary Material\cite{Booth_2021_SI}) starts with the linear in optical field equations of motion for the reduced density matrix. These are derived in the work of the \textit{Mukamel} group, and describe optical response of an electron-hole pair conserving many-body model, characterized by a set of occupied and unoccupied molecular orbitals \cite{mukamel1995, Axt}. 

A closed system of the kinetic equation has been derived by means of the time-dependent Hartree-Fock technique \cite{Takahashi, Yokojima} solving the hierarchy problem.\cite{kira} Unlike the conventional two-band semiconductor Bloch equations, this approach allows Coulomb coupling between subbands as well as local-fields effects to be taken into account, and is suitable for both Wannier-Mott and Frenkel excitons. This method has been successfully applied to finite and anisotropic structures such as conjugated polymers \cite{Takahashi}, and semiconductor nanocrystals.\cite{Yokojima}

In the linear optical regime, the optical properties of the system are determined by the kinetics of the non-diagonal density matrix elements, assuming that the diagonal elements (populations) are constant \cite{Haug}. The diagonal elements in this work are given by the ground state populations of the molecular orbitals. The resulting equation of motion can be written in a form reminiscent the semiconductor Bloch equations showing explicitly the common terms:

\begin{multline}
	i \hbar\frac{d}{dt} \rho_{ji,\mathbf{k}}^{vc} = (\epsilon_{\mathbf{k},j}^c-\epsilon_{\mathbf{k},i}^v - i\gamma) \rho_{ji,\mathbf{k}}^{vc} - d_{ij,\mathbf{k}}^{vc} \mathcal{E}(t)+\\
	\sum_{\mathbf{q} \neq \mathbf{k}} V_{|\mathbf{k}-\mathbf{q}|}^{ii|jj} \rho_{ji,\mathbf{q}}^{vc}+
	\sum_{\substack{\mathbf{q}, p \neq i \text{ or} \\ q \neq j}} \left( V_{|\mathbf{k}-\mathbf{q}|}^{pi|jq} - V_{|\mathbf{k}-\mathbf{q}|}^{pq|ji} \right) \rho_{qp,\mathbf{q}}^{vc}
	\label{sbe}
\end{multline}
where $\rho_{ij,\mathbf{k}}^{vc}$ is the non-diagonal element of the reduced density matrix for a pair of states with the wave vector $\mathbf{k}$ and band indices $i$ and $j$, $\epsilon_{\mathbf{k},j}^c$ and $\epsilon_{\mathbf{k},i}^v$ are the energies of electrons and holes respectively, represented by the quasi-particle energies from the GW computations, $d_{ij,\mathbf{k}}^{vc}$ is the dipole matrix element computed using DFT, $\mathcal{E}(t)$ is the external electromagnetic field, $V_{|\mathbf{k}-\mathbf{q}|}^{pi|jq}$ is the two-electron Coulomb potential\cite{Olsen} and $\gamma$ is a phenomenological de-phasing factor. This parameter can be considered as a rough approximation to the self-energy of all scattering processes determining broadening of spectral characteristics such as electron-electron, electron-phonon scatterings etc. which is generally a wave-vector, temperature, and energy-dependent function. In this work it is set to 0.1 eV, to match the broadening used in the BSE calculations (see the Supplementary Material\cite{Booth_2021_SI}). 

Eq. (\ref{sbe}) differs from the conventional semiconductor Bloch equations by the last sum which is responsible for the local-field effects, and causes coupling between microscopic polarizations for different pairs of energy bands. Its form is similar to the density matrix formulation of the TDDFT method, but the way the exchange-correlation kernel and orbital energies are defined is different. 

Since we are interested in the stationary optical spectra in this work, Eq. (\ref{sbe}) has been transformed into a linear algebra problem using a Fourier transform, and solved numerically (see Supplementary Material for more details\cite{Booth_2021_SI}). The first term in the second sum in Eq. (\ref{sbe}) corresponds to the direct Coulomb coupling, while the second one is the exchange energy. The numerical results show that the maximal contribution from the exchange term is five times smaller than the maximal direct term. The exchange term, unlike the direct Coulomb coupling, depends on the overlap of the wave functions of initial and final quantum states. The fact that it is small implies that the electron and hole wave functions do not overlap significantly meaning that the electron and hole form an interlayer exciton.

\subsection{Coulomb coupling}
The two-electron Coulomb integral in momentum space over the molecular orbitals reads: $\hat{V} = e^2\int d\mathbf{k} \tilde{\psi}_{pq}(-\mathbf{k}) g(\mathbf{k}) \tilde{\psi}_{nm}(\mathbf{k})$, where $g(\mathbf{k})$ is the Green's function of the Laplacian operator in momentum space and $\tilde{\psi}_{nm}(\mathbf{k}) = \mathcal{F} \left[  \hat{\psi}_{n}^*(\mathbf{r}) \hat{\psi}_{m}(\mathbf{r}) \right]$ is the Fourier transform of a product of field operators, $e$ is the elementary charge and $V$ is volume. Note that the integration is performed over a three-dimensional vector space, while in real space the two-electron integrals are six-dimensional. The considered material is characterized by anisotropic dielectric properties that change dramatically for the directions along and perpendicular to atomic chains. For such media, assuming axial symmetry, the Green's function for the Laplacian operator is defined by the Fourier-transformed Poisson equation: $\left( \epsilon_{\perp} \mathbf{k}_{\perp}^2 +  \epsilon_{\parallel} k_{\parallel}^2 \right) g(\mathbf{k}_{\perp}, k_{\parallel})= 4\pi$.\cite{Meckbach2018} For crystalline structures it is convenient to represent the momentum vector as a sum of the reciprocal lattice vector, $\mathbf{G}$, and the wave vector bounded within the first Brilloiun zone, $\mathbf{q}$: $\mathbf{k}_{\perp} = \mathbf{q}_{\perp}+\mathbf{G}_{\perp}$ and $k_{\parallel}=q_{\parallel} + G_{\parallel}$. The Coulomb potential operator in the representation
of bulk semiconductor states reads:

\begin{equation}
	\begin{split}
		\hat{V} = \sum_{\mathbf{G}_{\perp}, G_{\parallel}}  &B_{-\mathbf{q}_{\perp},  -q_{\parallel}}^{\lambda', \lambda}(-\mathbf{G}_{\perp},  -G_{\parallel}) B_{\mathbf{q}_{\perp},  q_{\parallel}}^{\nu', \nu}(\mathbf{G}_{\perp},  G_{\parallel}) \times \\
		&\frac{4 \pi e^2}{\epsilon_{\perp} (\mathbf{q}_{\perp}+\mathbf{G}_{\perp})^2 +  \epsilon_{\parallel} (q_{\parallel} + G_{\parallel})^2}    
	\end{split}
\label{f_pot}
\end{equation}
where $\mathbf{G}_{\perp}$ and $G_{\parallel}$ are the projections of the reciprocal lattice vectors in the direction along the atomic chains and on the plane perpendicular to the chains respectively, $\mathbf{q}_{\perp}$ and $q_{\parallel}$ are the projections of the wave vectors, confined within the first Brillouin zone, on the direction along atomic chains and on the plane perpendicular to the chains respectively and

\begin{multline}
	B_{\substack{\mathbf{k}_{\perp},  k_{\parallel},\\ \mathbf{q}_{\perp},  q_{\parallel}}}^{\lambda', \lambda}\left(\mathbf{G}_{\perp},  G_{\parallel} \right) = \int\limits_{\Omega} d\mathbf{r}_{\perp} dr_{\parallel} e^{i(\mathbf{G}_{\perp}\mathbf{r}_{\perp}  + G_{\parallel}r_{\parallel})} \times \\ u_{\lambda, \mathbf{k}_{\perp} + \mathbf{q}_{\perp},  k_{\parallel}+q_{\parallel}}^*(\mathbf{r}_{\perp}, r_{\parallel})u_{\lambda', \mathbf{k}_{\perp},  k_{\parallel}}(\mathbf{r}_{\perp}, r_{\parallel}) 
\end{multline}
where $\Omega$ is the unit cell volume, $u_{\lambda', \mathbf{q}_{\perp},  q_{\parallel}}(\mathbf{r}_{\perp}, r_{\parallel})$ are the periodic Bloch functions.

In the long wavelength limit where $\mathbf{G}_{\perp}=0$, $G_{\parallel}=0$  and $B_{\substack{\mathbf{k}_{\perp},  k_{\parallel},\\ \mathbf{q}_{\perp},  q_{\parallel}}}^{\lambda', \lambda}\left(\mathbf{G}_{\perp} =0 ,  G_{\parallel} = 0 \right) = \delta_{\lambda', \lambda}  \delta_{ \mathbf{q}_{\perp}, 0}  \delta_{ q_{\parallel}, 0} $ , Eq. (\ref{f_pot}) is further simplified as:

\begin{equation}
	\hat{V}_{\mathbf{q} = \{\mathbf{q}_{\perp}, q_{\parallel}\} } = 4 \pi e^2  \frac{1}{\epsilon_{\perp} \mathbf{q}_{\perp}^2 +  \epsilon_{\parallel} q_{\parallel}^2}
	\label{V0}
\end{equation}

The long wavelength limit is usually used in the semiconductor Bloch equations for Wannier-Mott excitons with large effective Bohr radii. In the semiconductor Bloch equations used in this study (see Supplementary Material\cite{Booth_2021_SI}), a long wavelength potential is used. For the BiSBr crystals this approximation can be very inaccurate since the situation when electron and hole are confined on the neighboring chains within the same primitive cell is possible and, as a result, the local field effects become important. 

Therefore, we need to take into account terms for which $\mathbf{G}_{\perp} > 0$. For BiSBr the dimensions of the Brillouin zone in the directions perpendicular to the atomic chains are much smaller compared to the size along the chains. This implies $|G_{\parallel}| >> |\mathbf{G}_{\perp}|$. As a result of this observation we neglect all terms for which $G_{\parallel} \neq 0$. Note that $\mathbf{G}_{\perp}$ can be comparable to $q_{\parallel}$. Eq. (\ref{VI}) can be further simplified considering optical transitions only around the center of the Brillouin zone neglecting weak dependence of $B_{\substack{\mathbf{k}_{\perp},  k_{\parallel},\\ \mathbf{q}_{\perp},  q_{\parallel}}}^{\nu', \nu}$ on $\mathbf{k}_{\perp},  k_{\parallel}, \mathbf{q}_{\perp}$ and $q_{\parallel}$:

\begin{equation}
	\hat{V}_{\mathbf{q} = \{\mathbf{q}_{\perp}, q_{\parallel}\} }^{\lambda, \lambda', \nu, \nu'} = 4 \pi e^2 \sum_{\mathbf{G}_{\perp}} \frac{B_0^{\lambda', \lambda}(-\mathbf{G}_{\perp}) B_0^{\nu', \nu}(\mathbf{G}_{\perp}) }{\epsilon_{\perp} (\mathbf{q}_{\perp}+\mathbf{G}_{\perp})^2 +  \epsilon_{\parallel} q_{\parallel}^2}
	\label{VI}
\end{equation}

Computing $B_0^{\nu', \nu}(\mathbf{G}_{\perp})$ explicitly from the DFT Kohn-Sham wave function using a Fast-Fourier transform confirms that the largest contributions have four reciprocal lattice vectors: $$\mathbf{G}_{\perp} = \{ [0, 2\pi / a], [0, -2\pi / a],[2\pi / c, 0],[-2\pi / c, 0]  \}.$$ All other components are neglected. The potential of (\ref{VI}) corresponds to the second term on the second line of Eq. (\ref{sbe}). 

In this work we are interested in the absorption around the band edges. For that reason, we compute the semiconductor Bloch equations for the system of the two lowest conduction bands and six topmost valence bands. The values of the static dielectric constants were computed using the GW Approximation: $\epsilon_{\parallel}=9.24$ and   $\epsilon_{\perp}=5.21$. It has been shown in Ref. \cite{Ulrich2015} that a constant dielectric screening works well for predicting the exciton binding energy in the framework of TDDFT. 
\subsection{Ab Initio Calculations}
The crystal structure parameters of BiSBr were obtained from the literature\cite{BiSBr}. All \textit{ab initio} calculations performed using Projector Augmented Waves\cite{Blochl1994b} and the Vienna Ab Initio Simulation Package (VASP)\cite{Kresse1996}. The DFT functional used in all calculations was the GGA functional of Perdew, Burke and Ernzerhof\cite{Perdew1996}, and all \textit{ab initio} calculations on this structure included spin-orbit interactions. 

The hybrid functional used was the HSE06\cite{Heyd2003} functional, and the GW band structure and input data for the Bloch Equations  were calculated on a 6$\times$6$\times$6 Monkhorst-Pack\cite{Monkhorst1976} k-point mesh using a single-shot G$_{0}$W$_0$ approach\cite{Shishkin2006}. All TDDFT and BSE calculations used the Tetrahedron method for Brillouin Zone integration with Bl\"ochl corrections\cite{Bloechl1994}, except for the Hybrid-DFT calculations which used Methfessel and Paxton smearing\cite{Methfessel1989}. The TDDFT and BSE calculations presented were performed on 4$\times$6$\times$4 Monkhorst-Pack k-point grids using 96 bands, and 3$\times$4$\times$3 Monkhorst-Pack k-point grids using 144 bands respectively in the Tamm-Dancoff approximation\cite{Onida2002}. The finer sampling in the \textbf{b} direction reflects the anistopy of the BiSBr unit cell, in which \textbf{a} = 8.47 \AA, \textbf{b} = 4.09 \AA and \textbf{c} = 10.58 \AA. For the \textit{ab initio} optical calculations the broadening applied (corresponding to the CSHIFT parameter in VASP) was 0.1 eV.

The eigenvalues and wavefunctions used as input to the SBE calculations used the GW Approximation approach implemented in VASP\cite{Shishkin2006} calculated on a 6$\times$6$\times$6 Monkhorst-Pack k-point mesh using the same input parameters and functional as that for the BSE calculations.

\section{Results and Discussion}
Given the aforementioned pseudo-one dimensional electronic structure of BiSBr and thus considerable confinement of excitons expected, the electronic band gap must differ significantly from the optical gap. That is the confinement induced stability will lower the energies of neutral excitations (given by the optical gap) compared to charged excitations (reflected in the electronic band gap). This stability will result in an optical gap which is lower than the band gap, and for low dimensional electronic structures, significantly so. 

Table \ref{BGs} lists the electronic band gaps of BiSBr computed using DFT, Hybrid DFT using the HSE06\cite{Heyd2003} functional, and the GW Approximation, compared with the experimental optical gap. The DFT and GW band gap values appear to roughly straddle the experimental optical gap of 2.01 eV; the GW calculation overshoots the experimental value by approximately 270 meV, while the DFT band gap is 0.65 eV lower. Attempting to correct the DFT gap by using the HSE06 hybrid functional instead of the PBE96\cite{Perdew1996} functional in the DFT  calculations gives a value of 1.927 eV, which brings it much closer to the experimental optical gap.
\begin{table}[!h]
\begin{center}
\begin{tabular}{c|c}\toprule
Technique& Gap (eV)\\
\hline
DFT& 1.36\\
HSE06& 1.93\\
GW& 2.28\\
Experiment& 2.01\\
\hline
\end{tabular}
\end{center}
\caption{Electronic band gaps determined from DFT, Hybrid DFT using the HSE06 functional, and the GW Approximation compared to the experimental optical gap.}
\label{BGs}
\end{table}

\begin{figure*}[!ht]
    \centering
    \subfigure[]{\includegraphics[width=0.66\columnwidth]{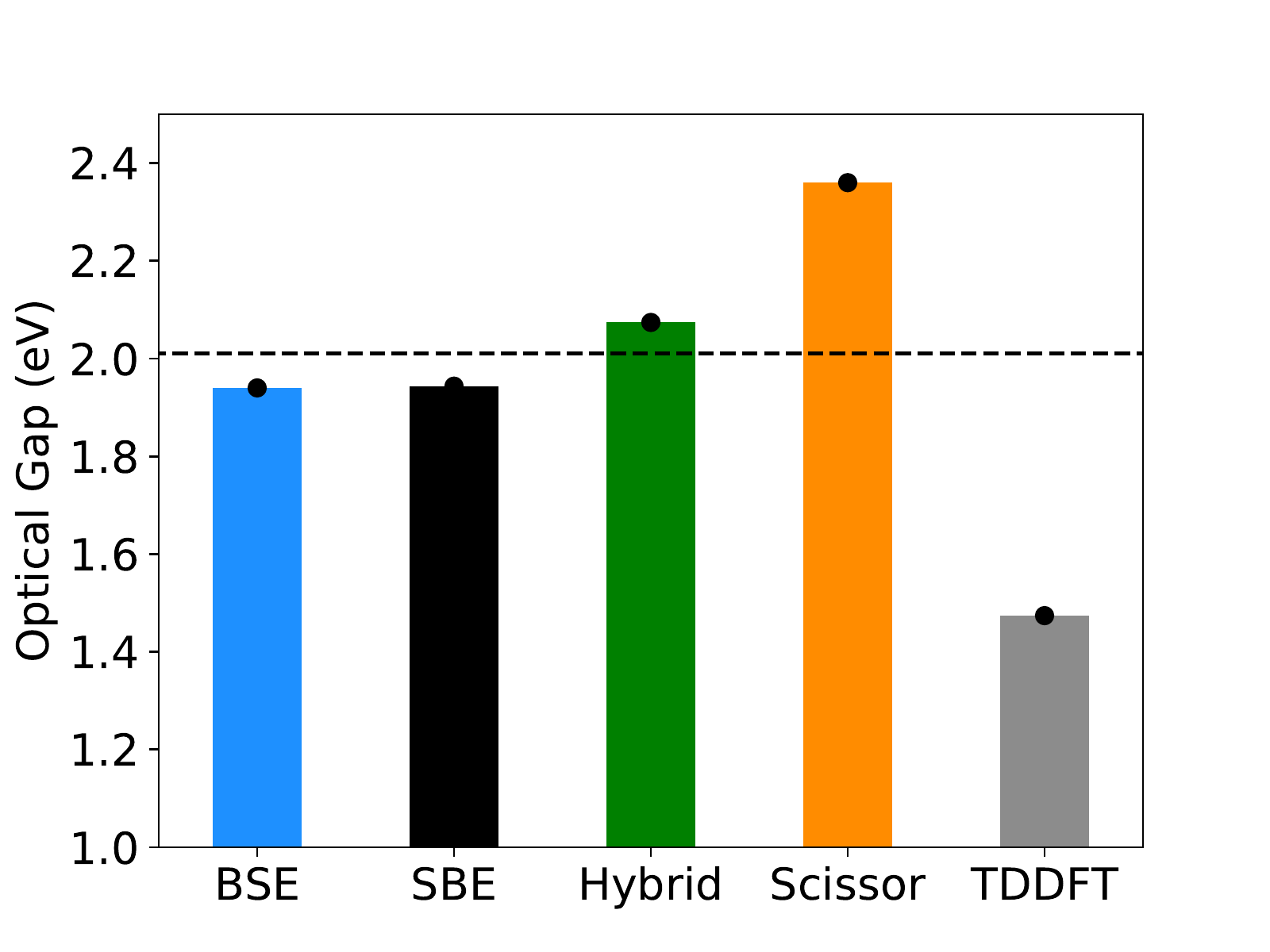}}
    \subfigure[]{\includegraphics[width=0.66\columnwidth]{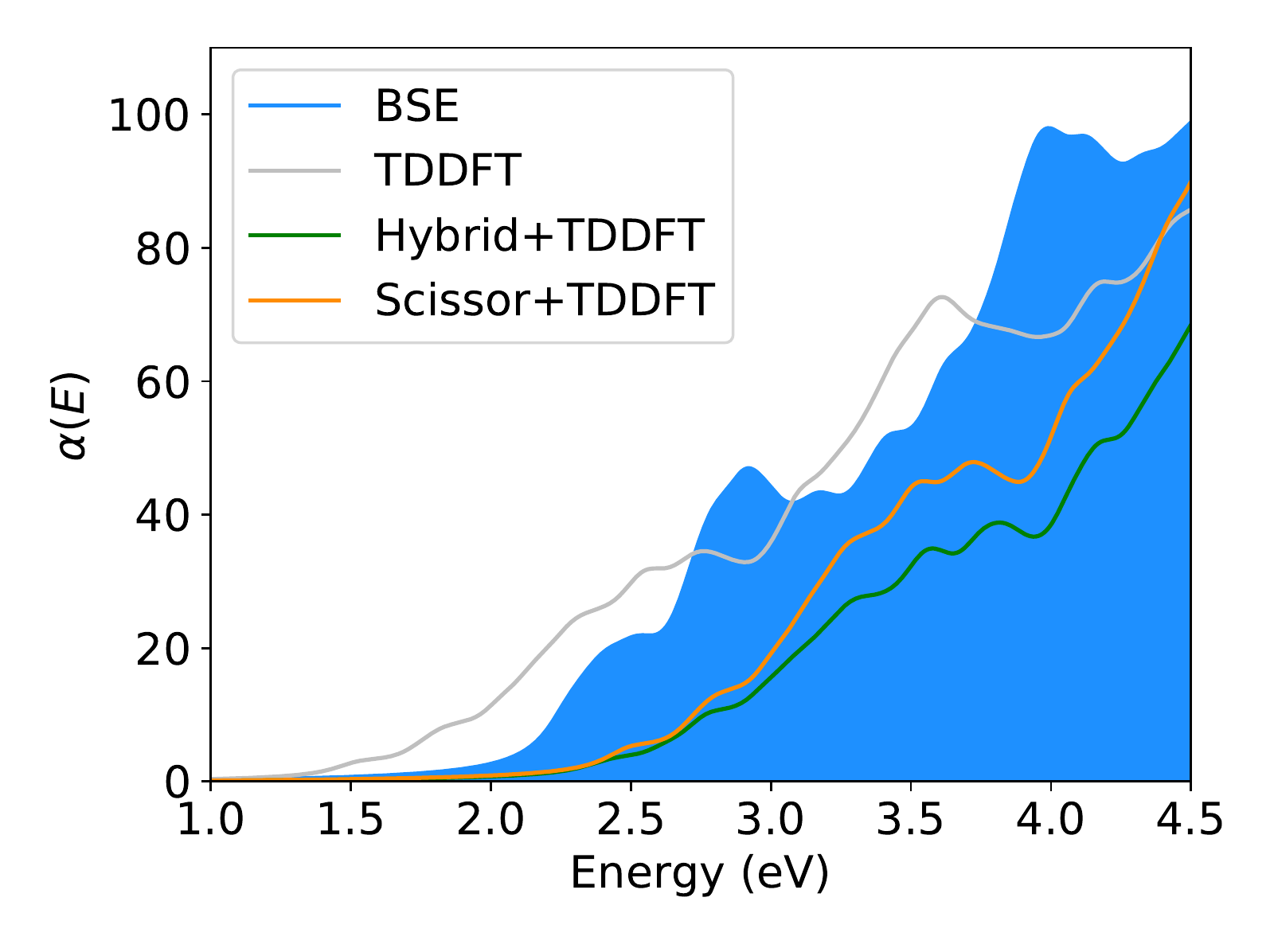}}
    \subfigure[]{\includegraphics[width=0.66\columnwidth]{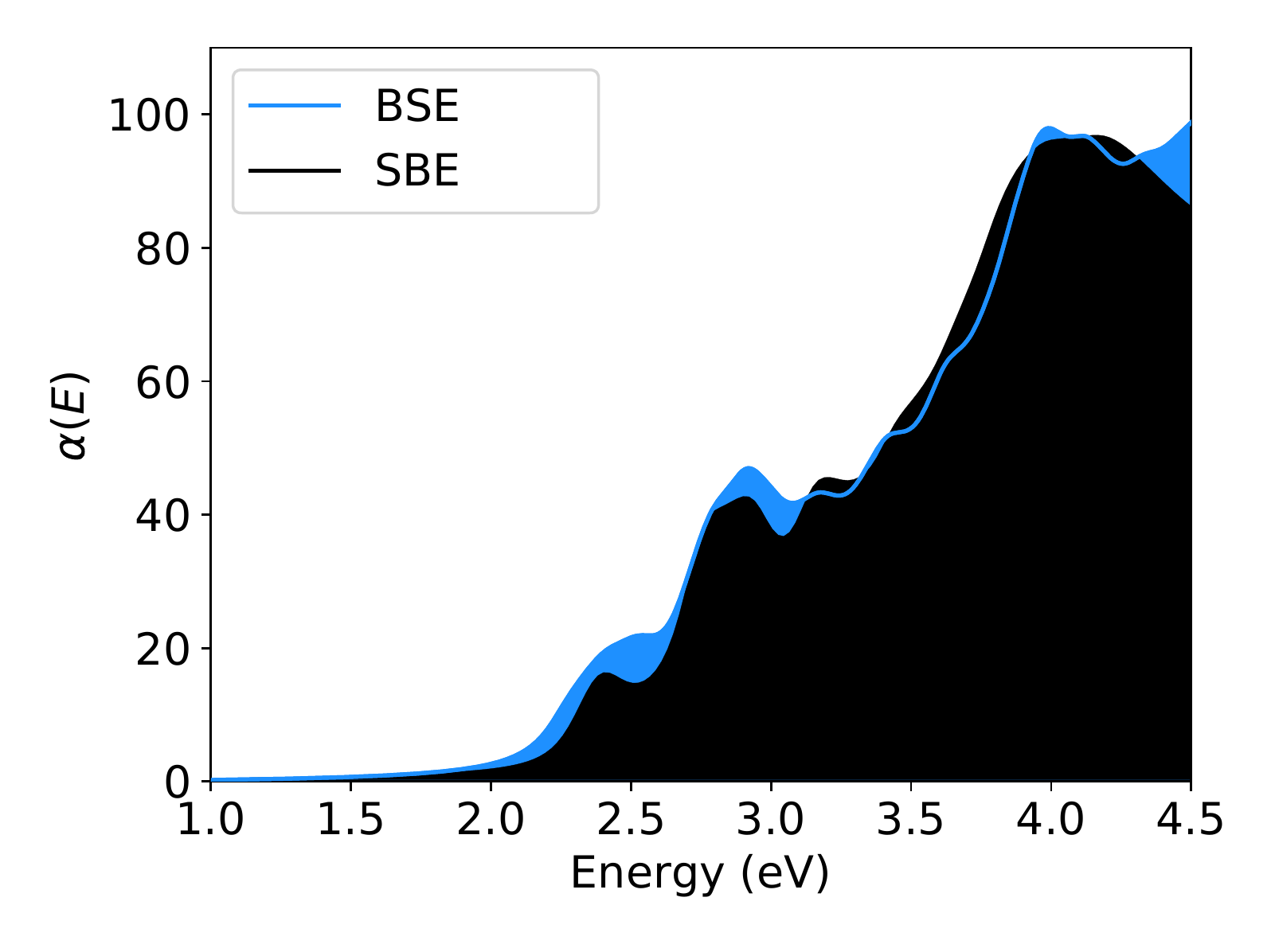}}
    \subfigure[]{\includegraphics[width=0.66\columnwidth]{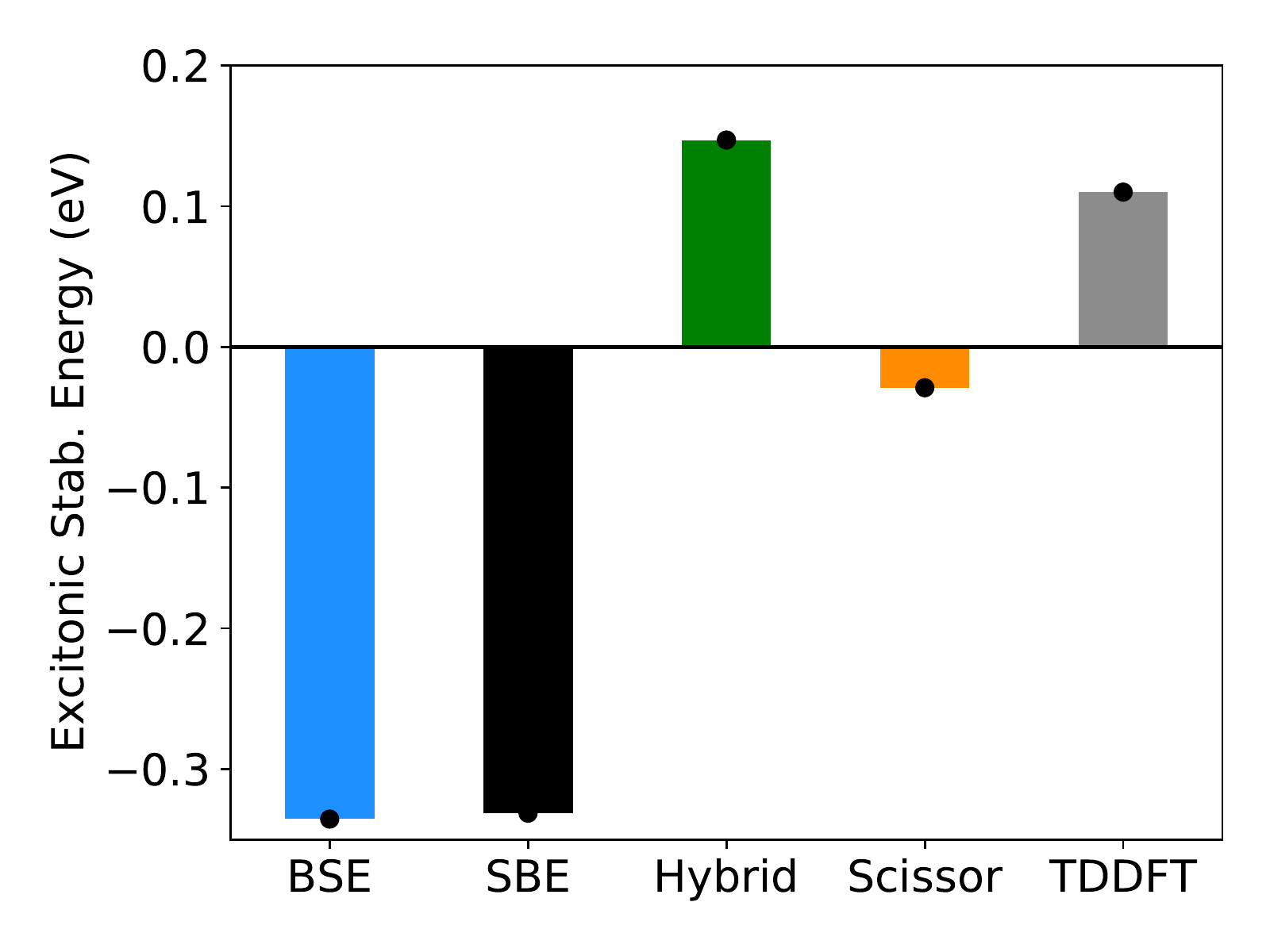}}
    \subfigure[]{\includegraphics[width=0.68\columnwidth]{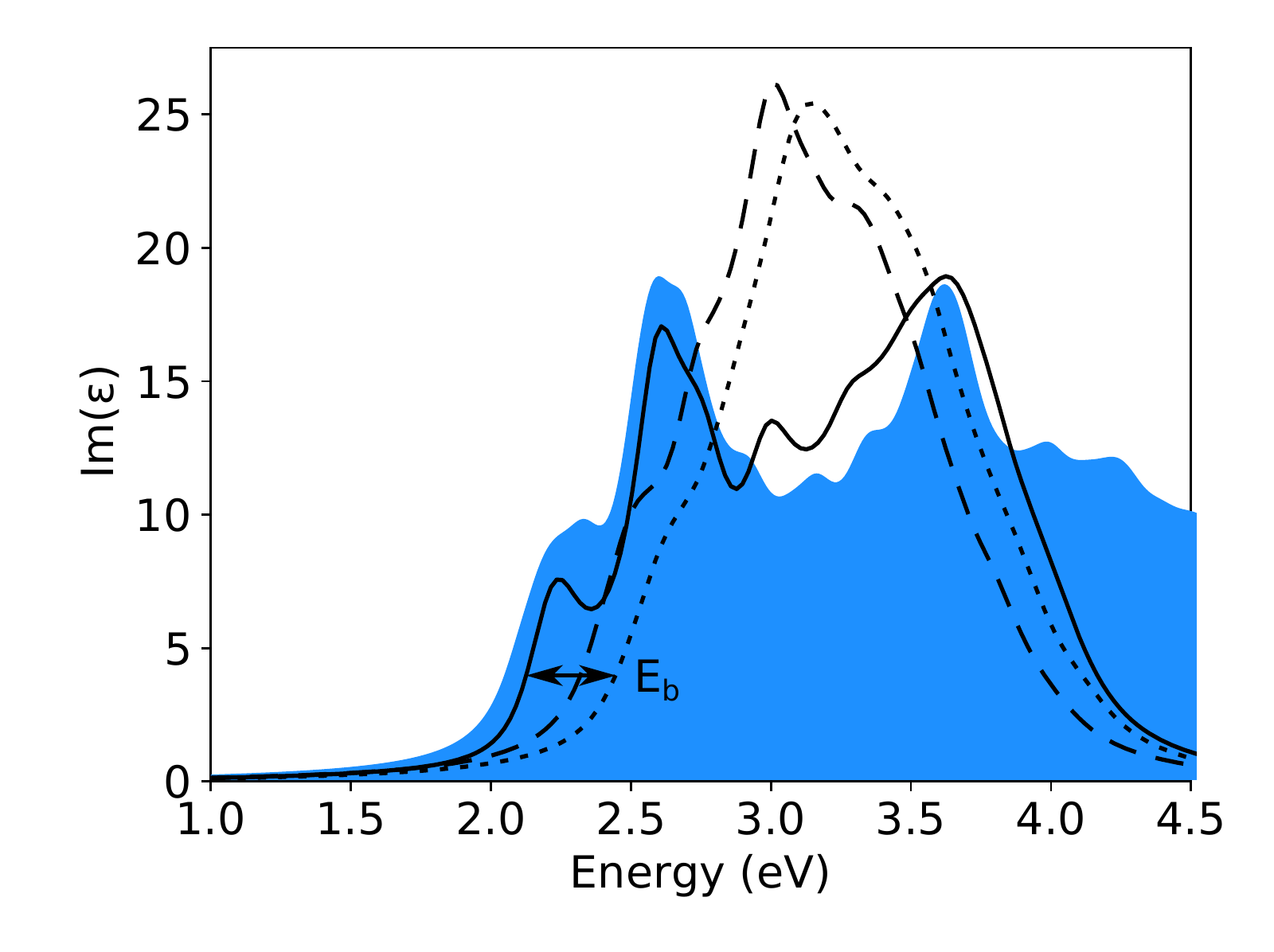}}
    \caption{(Color Online) Optical properties of BiSBr calculated using various techniques described in the text. a) Optical gaps computed using BSE, SBE, TDDFT, TDDFT with a scissor operator, and TDDFT using the HSE06 functional, with the experimental gap plotted as a dashed line, b) corresponding simulated absorption spectra not including the SBE approach c) absorption spectrum of SBE compared with BSE, d) Excitonic stabilisation energies of the five approaches, e) SBE imaginary part of the dielectric matrix for the direction along atomic chains computed with (solid line) and without (dashed line) local field effects, and with free particles (dotted line) compared to BSE (shaded blue curve). Comparing the free particle spectrum and excitonic spectrum gives an estimate of the exciton binding energy: $E_b = 305$ meV.}
    \label{Abs}
\end{figure*}

The large discrepancy between the DFT and GW band gaps most likely has its origin in the fact that DFT is overestimating the screening in comparison to the GW method, and that the electrons are interacting much more strongly than DFT indicates. This reduction in screening in the GW calculations will result in conduction electrons interacting more strongly with the valence band, thus increasing their energies. 

The band gap data therefore suggest that of the three approaches, only the GW gap has the ``wiggle room" to give an optical gap in line with experiment once excitonic effects in the case of BSE calculations, or quasiparticle effects in the case of TDDFT are included. Given that both DFT and HSE06 predict band gaps lower in energy than the optical gap it is unlikely that including quasiparticle effects using TDDFT will result in better agreement with experiment, while also exhibiting considerable excitonic stabilisation.

However, we can try to improve on this. By using a scissor operator on the DFT data to align its band gap with that from GW and performing TDDFT calculations there may be some hope of correcting the DFT data. Likewise, we can use the GW eigenvalues and dipole matrix elements as input into semiconductor Bloch equations in an attempt to get accurate optical data. Both of these approaches will be less computationally expensive than BSE, and in the case of semiconductor Bloch equations, significantly so.

Figure \ref{Abs} presents the optical data of BiSBr calculated using five different approaches: the Bethe-Salpeter method (BSE), the semiconductor Bloch Equation approach (SBE), Time-Dependent Density Functional Theory (TDDFT), TDDFT combined with a hybrid DFT functional (again using the HSE06\cite{Heyd2003} functional, henceforth simply hybrid-TDDFT) in an attempt to start from a more accurate DFT band gap, and TDDFT in which a scissor operator has been applied to the conduction bands also in an attempt to correct the DFT band gap. The value of the scissor operator was chosen to align the DFT band gap with that from GW calculations. 

The data of Figure \ref{Abs}a plots the optical gaps resulting from the five approaches determined from Tauc plots\cite{Tauc1966} in order to be consistent with the experimental value (see the Supplementary Material for an example\cite{Booth_2021_SI}), with the experimental gap represented by a dashed line. Of the five approaches the BSE, Bloch Equation and hybrid-TDDFT come the closest to reproducing the experimental gap of 2.01 eV, with the TDDFT+scissor approach and plain TDDFT significantly over- and underestimating the gap respectively. 

However while the optical gap from the hybrid-TDDFT approach appears reasonable, the optical absorption spectra of Figure \ref{Abs}b and the exciton stabilization energies plotted in Figure \ref{Abs}f suggest otherwise. Comparing TDDFT and BSE (Fig \ref{Abs}b, gray trace) we see that the TDDFT spectrum is shifted to lower energy, indicating a smaller gap, and that the peaks in the spectrum do not match those of the BSE approach, indicating a significant difference in the energy distribution of the optical transitions and the oscillator strengths.

The TDDFT approach combined with the scissor operator (Figure \ref{Abs}b, orange trace) shifts the absorption to higher energy which it was designed to do, but in doing so overshoots the BSE spectrum and again the peaks do not line up. The hybrid approach (green trace) fares similarly; the spectrum is shifted to higher energy, slightly overshooting the experimental optical gap, and peak shapes differ significantly. Neither TDDFT-derived spectrum exhibits any significant red-shift.

However, the Bloch equation approach (Fig. \ref{Abs}c) returns an absorption spectrum which is almost identical to the BSE data. This tells us that the BSE and Bloch Equation approaches are consistent with each other, but not the hybrid-TDDFT data.

The question then becomes: is it the hybrid-TDDFT data or the BSE and SBE data that is correct? Figure \ref{Abs}d answers this by plotting the excitonic stabilization energies; the red shift of the absorption spectrum compared to the band gap, computed from the differences between the band gaps, and the optical gaps of the five approaches. Note that this is a different quantity to the exciton binding energy calculated in Figure \ref{Abs}e. 

The data clearly show that of the three techniques which get the closest to the experimental optical gap, only the BSE and Bloch Equation approach predict a considerable stabilization energy in line with what is expected from the pseudo one-dimensional BiSBr structure. In fact the hybrid approach predicts unstable excitons (positive energy).

As discussed above, the pseudo one-dimensional nature of the electronic structure of BiSBr will result in confinement of excitons, which stabilizes them against dissociation. Combining this with the well studied suitability of metal halide materials to excitonic applications\cite{Brandt2015,Fu2019,Shamsi2019}, we would expect the optical gap to be significantly lower than the electronic band gap due to excitonic stabilization. The TDDFT data is inconsistent with this, while the BSE and SBE data are entirely in line with this expectation, while also providing good agreement with the experimental optical gap.

Therefore, from knowledge of the crystal and electronic structures it is clear that the agreement of the optical gap of the hybrid-TDDFT data with experiment is coincidence. The blue shift of the band gap due to the exact exchange introduced by the hybrid functional coincidentally shifts the optical gap to a value close to experiment. The reason for the failure of the TDDFT technique for this material is that it does not contain the required electron-hole interactions\cite{Onida2002} (see the Supplementary Material\cite{Booth_2021_SI}), and therefore cannot reproduce the stabilization of optical excitations due to these interactions, which are considerable in pseudo-one-dimensional materials like BiSBr.

Thus we see that the BSE and Bloch Equation approaches predict an optical gap within 3.5 \% of experiment, and also predict excitonic stabilisation energies of $>$ 0.3 eV, in line with what would be expected from a metal halide with a pseudo one-dimensional electronic structure. The data from TDDFT either fares poorly at reproducing the optical gap, or predicts positive  excitonic stabilization energy (i.e. a repulsive interaction). 

This is a remarkable result, as the computational requirements of the BSE and Bloch equation approaches occupy opposite ends of the spectrum. The BSE calculations use $\sim$ 6,000 CPU hours, while the SBE approach uses just 0.14 CPU hours. Despite this difference, the techniques agree very closely on the most significant aspects of the optical properties: the absorption spectra are almost identical.

\begin{figure}[!h]
	\centering
	\subfigure[]{\includegraphics[height=0.47\linewidth]{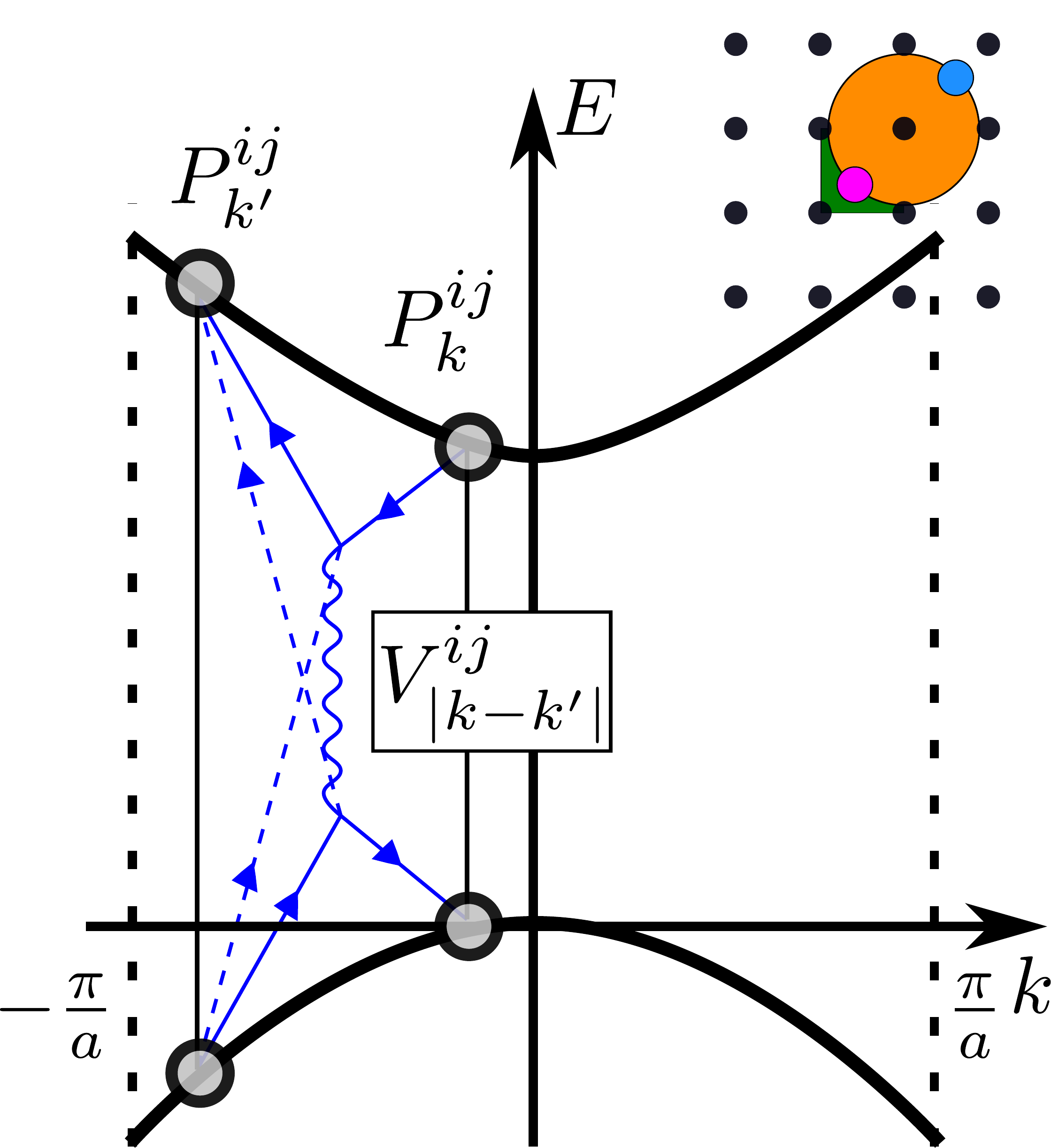}}
	\subfigure[]{\includegraphics[height=0.47\linewidth]{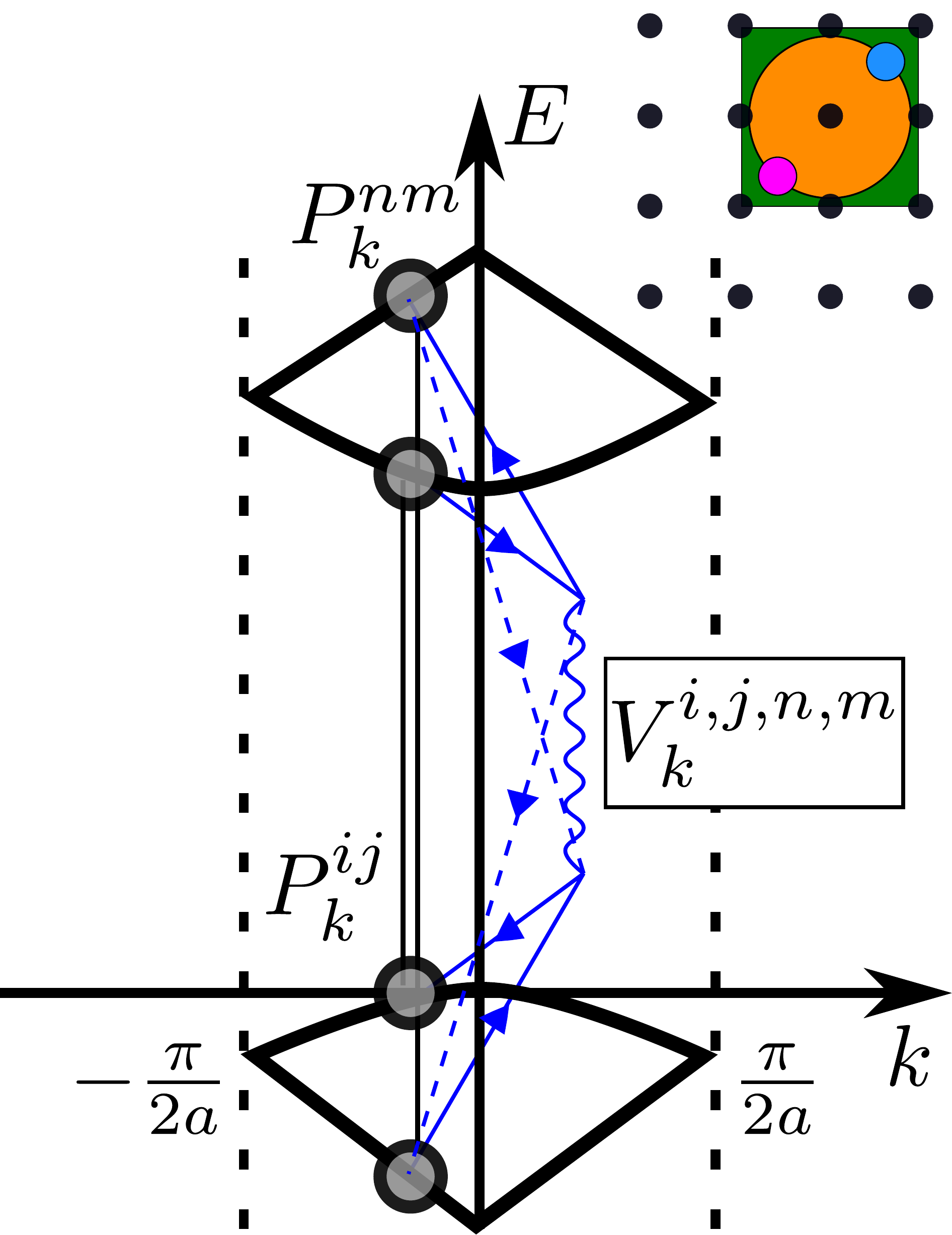}}
	\caption{(Color Online) a) Coulomb couping $V^{ij}_{|k-k'|}$ between microscopic polarization operators $P^{ij}_{k}$ and $P^{ij}_{k'}$ for Wannier-Mott excitons with Bohr radii much larger than a lattice constant couples electrons and holes with different quasi-momenta belonging to one pair of energy bands. b) Coulomb coupling for the Frenkel excitons that couples microscopic polarizations for different pairs of bands.}
	\label{fig:folding}
\end{figure}

The advantage of the SBE approach its efficiency in describing Wannier-Mott excitons with long range Coulomb interactions in an isotropic medium. This efficiency is achieved by reducing the number of degrees of freedom due to symmetry considerations. In this work we have achieved better performance  by reducing the number of bands participating in the optical response and wave vectors in the Coulomb coupling terms written in the momentum space representation, taking into account particularities of the band structure of BiSBr. 

For other systems, for instance in the case of III-V bulk semiconductors an axial approximation for the Coulomb potential and band structure is employed considering only the lowest conduction band and highest valence band. Unfortunately this approach is inaccurate in the case of BiSBr due to its anisotropic properties and strong exciton localization in the direction perpendicular to the atomic chains. Another way to reduce the computational cost is by applying a simplified dielectric screening model such as the static screened-exchange Coulomb-hole approximation\cite{Erben2018} or a plasmon-pole model with or without free parameters\cite{Engel1993}. 

In general though, the main advantage of using SBE over BSE are the time-dependent features of the former that allow access of the dynamical optical response and non-linear optical characteristics.

In Fig. \ref{Abs}e we compare the imaginary part of the dielectric constant computed with SBE approach and semiconductor Bloch equations and compare the exciton binding energies with and without local field effects (the exciton binding energy is the difference between the energy of the absorption edge without Coulomb interactions between the electrons and holes, and the exciton peak with these interactions included). The results indicate that good agreement between two methods can be achieved only if the local fields effect are taken into account (see Supplementary Material\cite{Booth_2021_SI}). Without these the exciton binding energy is significantly underestimated; on the order of $\sim$ 10 meV. 

However, taking the local field effects into account leads to a huge exciton binding energy ($\sim$ 300 meV) which implies that the exciton wave function is localized within the unit cell and is represented by a linear combination of the bulk crystal states having not only different wave vectors, but also different band indices. 

The contribution of the local field effects is negligible in most III-V semiconductors, where the exciton Bohr radius is much larger than the lattice constant (Wannier-Mott excitons). However, they are significant in perovskites and van-der-Waals heterostructures where the lattice constant in one direction is comparable to the Bohr radius. This can occur due to small dielectric screening on the one hand, or larger spacing between atomic layers (or chains in BiSBr) on the other hand. 

The relationship between the localization of the exciton wave function and local field effects can be illustrated by considering a simple rectangular lattice whose conduction and valence bands are shown schematically in Fig. \ref{fig:folding}a. The Coulomb potential in the convectional semiconductor Bloch equations couples states with different momenta, but is diagonal in the band indices. If we keep the same crystalline structure and double the lattice constant, the Brillouin zone halves, and the electronic bands fold over as per Fig. \ref{fig:folding}b. 

As a result, Coulomb coupling between two wave vectors in the first case becomes coupling between different bands in the second case. Thus, if we want to represent the wave function of the excitons with large binding energy as a linear combination of bulk semiconductor wave functions, it is not enough to use wave functions with different wave vectors for a pair of bands; several conduction and valence bands must be used.

\section{Conclusions}
To summarize, we find that unsurprisingly the computationally expensive, but theoretically rigorous GW-Bethe-Salpeter approach to computing optical properties produces predicts an optical gap in excellent agreement with experiment for the metal halide BiSBr, while TDDFT calculations did not.

However, combining the \textit{ab initio} approach with the semiconductor Bloch equation formalism appears to be a viable low-resource substitute for BSE. The data is in good agreement with the experimental optical gap (1.94 eV vs 2.01 eV, a difference of $\sim$ 3.5\%), and also agrees particularly well with the optical gap, excitonic stabilization energy and the absorption spectrum of the BSE data. This agreement comes with a significantly reduced computational cost, making the Bloch Equation approach a good candidate for theoretical spectroscopy on systems which are computationally out of reach of BSE calculations.

\section{Acknowledgments}
The authors acknowledge the support of the ARC Centre of Excellence in Exciton Science (CE170100026). This work was supported by computational resources provided by the Australian Government through the National Computational Infrastructure and the Pawsey Supercomputer Centre.

\section{Supplementary Material} Supplementary material which presents the background theory for the TDDFT and BSE approaches, and provides a detailed description of the SBE approach is available. 

\bibliography{C:/Bibliography/library}

\begin{thebibliography}{44}%
\makeatletter
\providecommand \@ifxundefined [1]{%
 \@ifx{#1\undefined}
}%
\providecommand \@ifnum [1]{%
 \ifnum #1\expandafter \@firstoftwo
 \else \expandafter \@secondoftwo
 \fi
}%
\providecommand \@ifx [1]{%
 \ifx #1\expandafter \@firstoftwo
 \else \expandafter \@secondoftwo
 \fi
}%
\providecommand \natexlab [1]{#1}%
\providecommand \enquote  [1]{``#1''}%
\providecommand \bibnamefont  [1]{#1}%
\providecommand \bibfnamefont [1]{#1}%
\providecommand \citenamefont [1]{#1}%
\providecommand \href@noop [0]{\@secondoftwo}%
\providecommand \href [0]{\begingroup \@sanitize@url \@href}%
\providecommand \@href[1]{\@@startlink{#1}\@@href}%
\providecommand \@@href[1]{\endgroup#1\@@endlink}%
\providecommand \@sanitize@url [0]{\catcode `\\12\catcode `\$12\catcode
  `\&12\catcode `\#12\catcode `\^12\catcode `\_12\catcode `\%12\relax}%
\providecommand \@@startlink[1]{}%
\providecommand \@@endlink[0]{}%
\providecommand \url  [0]{\begingroup\@sanitize@url \@url }%
\providecommand \@url [1]{\endgroup\@href {#1}{\urlprefix }}%
\providecommand \urlprefix  [0]{URL }%
\providecommand \Eprint [0]{\href }%
\providecommand \doibase [0]{http://dx.doi.org/}%
\providecommand \selectlanguage [0]{\@gobble}%
\providecommand \bibinfo  [0]{\@secondoftwo}%
\providecommand \bibfield  [0]{\@secondoftwo}%
\providecommand \translation [1]{[#1]}%
\providecommand \BibitemOpen [0]{}%
\providecommand \bibitemStop [0]{}%
\providecommand \bibitemNoStop [0]{.\EOS\space}%
\providecommand \EOS [0]{\spacefactor3000\relax}%
\providecommand \BibitemShut  [1]{\csname bibitem#1\endcsname}%
\let\auto@bib@innerbib\@empty
\bibitem [{\citenamefont {Stranks}\ \emph {et~al.}(2013)\citenamefont
  {Stranks}, \citenamefont {Eperon}, \citenamefont {Grancini}, \citenamefont
  {Menelaou}, \citenamefont {Alcocer}, \citenamefont {Leijtens}, \citenamefont
  {Herz}, \citenamefont {Petrozza},\ and\ \citenamefont
  {Snaith}}]{Stranks2013}%
  \BibitemOpen
  \bibfield  {author} {\bibinfo {author} {\bibfnamefont {S.~D.}\ \bibnamefont
  {Stranks}}, \bibinfo {author} {\bibfnamefont {G.~E.}\ \bibnamefont {Eperon}},
  \bibinfo {author} {\bibfnamefont {G.}~\bibnamefont {Grancini}}, \bibinfo
  {author} {\bibfnamefont {C.}~\bibnamefont {Menelaou}}, \bibinfo {author}
  {\bibfnamefont {M.~J.~P.}\ \bibnamefont {Alcocer}}, \bibinfo {author}
  {\bibfnamefont {T.}~\bibnamefont {Leijtens}}, \bibinfo {author}
  {\bibfnamefont {L.~M.}\ \bibnamefont {Herz}}, \bibinfo {author}
  {\bibfnamefont {A.}~\bibnamefont {Petrozza}}, \ and\ \bibinfo {author}
  {\bibfnamefont {H.~J.}\ \bibnamefont {Snaith}},\ }\href@noop {} {\bibfield
  {journal} {\bibinfo  {journal} {Science}\ }\textbf {\bibinfo {volume}
  {342}},\ \bibinfo {pages} {341} (\bibinfo {year} {2013})}\BibitemShut
  {NoStop}%
\bibitem [{\citenamefont {Yang}\ \emph
  {et~al.}(2015{\natexlab{a}})\citenamefont {Yang}, \citenamefont {Noh},
  \citenamefont {Jeon}, \citenamefont {Kim}, \citenamefont {Ryu}, \citenamefont
  {Seo},\ and\ \citenamefont {Seok}}]{Yang2015}%
  \BibitemOpen
  \bibfield  {author} {\bibinfo {author} {\bibfnamefont {W.~S.}\ \bibnamefont
  {Yang}}, \bibinfo {author} {\bibfnamefont {J.~H.}\ \bibnamefont {Noh}},
  \bibinfo {author} {\bibfnamefont {N.~J.}\ \bibnamefont {Jeon}}, \bibinfo
  {author} {\bibfnamefont {Y.~C.}\ \bibnamefont {Kim}}, \bibinfo {author}
  {\bibfnamefont {S.}~\bibnamefont {Ryu}}, \bibinfo {author} {\bibfnamefont
  {J.}~\bibnamefont {Seo}}, \ and\ \bibinfo {author} {\bibfnamefont {S.~I.}\
  \bibnamefont {Seok}},\ }\href {\doibase 10.1126/science.aaa9272} {\bibfield
  {journal} {\bibinfo  {journal} {Science}\ }\textbf {\bibinfo {volume}
  {348}},\ \bibinfo {pages} {1234} (\bibinfo {year}
  {2015}{\natexlab{a}})}\BibitemShut {NoStop}%
\bibitem [{\citenamefont {Brandt}\ \emph {et~al.}(2015)\citenamefont {Brandt},
  \citenamefont {Stevanovi{\'{c}}}, \citenamefont {Ginley},\ and\ \citenamefont
  {Buonassisi}}]{Brandt2015}%
  \BibitemOpen
  \bibfield  {author} {\bibinfo {author} {\bibfnamefont {R.~E.}\ \bibnamefont
  {Brandt}}, \bibinfo {author} {\bibfnamefont {V.}~\bibnamefont
  {Stevanovi{\'{c}}}}, \bibinfo {author} {\bibfnamefont {D.~S.}\ \bibnamefont
  {Ginley}}, \ and\ \bibinfo {author} {\bibfnamefont {T.}~\bibnamefont
  {Buonassisi}},\ }\href {\doibase 10.1557/mrc.2015.26} {\bibfield  {journal}
  {\bibinfo  {journal} {MRS Communications}\ }\textbf {\bibinfo {volume} {5}},\
  \bibinfo {pages} {265} (\bibinfo {year} {2015})}\BibitemShut {NoStop}%
\bibitem [{\citenamefont {Fu}\ \emph {et~al.}(2019)\citenamefont {Fu},
  \citenamefont {Zhu}, \citenamefont {Chen}, \citenamefont {Hautzinger},
  \citenamefont {Zhu},\ and\ \citenamefont {Jin}}]{Fu2019}%
  \BibitemOpen
  \bibfield  {author} {\bibinfo {author} {\bibfnamefont {Y.}~\bibnamefont
  {Fu}}, \bibinfo {author} {\bibfnamefont {H.}~\bibnamefont {Zhu}}, \bibinfo
  {author} {\bibfnamefont {J.}~\bibnamefont {Chen}}, \bibinfo {author}
  {\bibfnamefont {M.~P.}\ \bibnamefont {Hautzinger}}, \bibinfo {author}
  {\bibfnamefont {X.~Y.}\ \bibnamefont {Zhu}}, \ and\ \bibinfo {author}
  {\bibfnamefont {S.}~\bibnamefont {Jin}},\ }\href {\doibase
  10.1038/s41578-019-0080-9} {\bibfield  {journal} {\bibinfo  {journal} {Nature
  Reviews Materials}\ }\textbf {\bibinfo {volume} {4}},\ \bibinfo {pages} {169}
  (\bibinfo {year} {2019})}\BibitemShut {NoStop}%
\bibitem [{\citenamefont {Shamsi}\ \emph {et~al.}(2019)\citenamefont {Shamsi},
  \citenamefont {Urban}, \citenamefont {Imran}, \citenamefont {Trizio},\ and\
  \citenamefont {Manna}}]{Shamsi2019}%
  \BibitemOpen
  \bibfield  {author} {\bibinfo {author} {\bibfnamefont {J.}~\bibnamefont
  {Shamsi}}, \bibinfo {author} {\bibfnamefont {A.~S.}\ \bibnamefont {Urban}},
  \bibinfo {author} {\bibfnamefont {M.}~\bibnamefont {Imran}}, \bibinfo
  {author} {\bibfnamefont {L.~D.}\ \bibnamefont {Trizio}}, \ and\ \bibinfo
  {author} {\bibfnamefont {L.}~\bibnamefont {Manna}},\ }\href@noop {}
  {\bibfield  {journal} {\bibinfo  {journal} {Chemical Reviews}\ }\textbf
  {\bibinfo {volume} {119}},\ \bibinfo {pages} {3296} (\bibinfo {year}
  {2019})}\BibitemShut {NoStop}%
\bibitem [{\citenamefont {Burke}(2012)}]{Burke2012}%
  \BibitemOpen
  \bibfield  {author} {\bibinfo {author} {\bibfnamefont {K.}~\bibnamefont
  {Burke}},\ }\href {\doibase 10.1063/1.4704546} {\bibfield  {journal}
  {\bibinfo  {journal} {J. Chem. Phys.}\ }\textbf {\bibinfo {volume} {136}}
  (\bibinfo {year} {2012}),\ 10.1063/1.4704546}\BibitemShut {NoStop}%
\bibitem [{\citenamefont {Aryasetiawan}\ and\ \citenamefont
  {Gunnarsson}(1998)}]{Aryasetiawan1998}%
  \BibitemOpen
  \bibfield  {author} {\bibinfo {author} {\bibfnamefont {F.}~\bibnamefont
  {Aryasetiawan}}\ and\ \bibinfo {author} {\bibfnamefont {O.}~\bibnamefont
  {Gunnarsson}},\ }\href@noop {} {\bibfield  {journal} {\bibinfo  {journal}
  {Reports Prog. Phys.}\ }\textbf {\bibinfo {volume} {61}},\ \bibinfo {pages}
  {237} (\bibinfo {year} {1998})}\BibitemShut {NoStop}%
\bibitem [{\citenamefont {Onida}\ \emph {et~al.}(2002)\citenamefont {Onida},
  \citenamefont {Reining},\ and\ \citenamefont {Rubio}}]{Onida2002}%
  \BibitemOpen
  \bibfield  {author} {\bibinfo {author} {\bibfnamefont {G.}~\bibnamefont
  {Onida}}, \bibinfo {author} {\bibfnamefont {L.}~\bibnamefont {Reining}}, \
  and\ \bibinfo {author} {\bibfnamefont {A.}~\bibnamefont {Rubio}},\
  }\href@noop {} {\bibfield  {journal} {\bibinfo  {journal} {Reviews of Modern
  Physics}\ }\textbf {\bibinfo {volume} {74}},\ \bibinfo {pages} {601}
  (\bibinfo {year} {2002})}\BibitemShut {NoStop}%
\bibitem [{\citenamefont {Casida}\ and\ \citenamefont
  {Huix-Rotllant}(2012)}]{Casida2012}%
  \BibitemOpen
  \bibfield  {author} {\bibinfo {author} {\bibfnamefont {M.~E.}\ \bibnamefont
  {Casida}}\ and\ \bibinfo {author} {\bibfnamefont {M.}~\bibnamefont
  {Huix-Rotllant}},\ }\href {\doibase 10.1146/annurev-physchem-032511-143803}
  {\bibfield  {journal} {\bibinfo  {journal} {Annual Review of Physical
  Chemistry}\ }\textbf {\bibinfo {volume} {63}},\ \bibinfo {pages} {287}
  (\bibinfo {year} {2012})},\ \Eprint {http://arxiv.org/abs/1108.0611}
  {arXiv:1108.0611} \BibitemShut {NoStop}%
\bibitem [{\citenamefont {Reining}\ \emph {et~al.}(2002)\citenamefont
  {Reining}, \citenamefont {Olevano}, \citenamefont {Rubio},\ and\
  \citenamefont {Onida}}]{Reining2002}%
  \BibitemOpen
  \bibfield  {author} {\bibinfo {author} {\bibfnamefont {L.}~\bibnamefont
  {Reining}}, \bibinfo {author} {\bibfnamefont {V.}~\bibnamefont {Olevano}},
  \bibinfo {author} {\bibfnamefont {A.}~\bibnamefont {Rubio}}, \ and\ \bibinfo
  {author} {\bibfnamefont {G.}~\bibnamefont {Onida}},\ }\href {\doibase
  10.1103/PhysRevLett.88.066404} {\bibfield  {journal} {\bibinfo  {journal}
  {Phys. Rev. Lett.}\ }\textbf {\bibinfo {volume} {88}},\ \bibinfo {pages}
  {66404} (\bibinfo {year} {2002})}\BibitemShut {NoStop}%
\bibitem [{\citenamefont {Cunningham}\ \emph {et~al.}(2018)\citenamefont
  {Cunningham}, \citenamefont {Gr{\"{u}}ning}, \citenamefont {Azarhoosh},
  \citenamefont {Pashov},\ and\ \citenamefont {{Van
  Schilfgaarde}}}]{Cunningham2018}%
  \BibitemOpen
  \bibfield  {author} {\bibinfo {author} {\bibfnamefont {B.}~\bibnamefont
  {Cunningham}}, \bibinfo {author} {\bibfnamefont {M.}~\bibnamefont
  {Gr{\"{u}}ning}}, \bibinfo {author} {\bibfnamefont {P.}~\bibnamefont
  {Azarhoosh}}, \bibinfo {author} {\bibfnamefont {D.}~\bibnamefont {Pashov}}, \
  and\ \bibinfo {author} {\bibfnamefont {M.}~\bibnamefont {{Van
  Schilfgaarde}}},\ }\href {\doibase 10.1103/PhysRevMaterials.2.034603}
  {\bibfield  {journal} {\bibinfo  {journal} {Physical Review Materials}\
  }\textbf {\bibinfo {volume} {2}},\ \bibinfo {pages} {1} (\bibinfo {year}
  {2018})}\BibitemShut {NoStop}%
\bibitem [{\citenamefont {Haug}\ and\ \citenamefont
  {Koch}(2004{\natexlab{a}})}]{Haug2004_BEs}%
  \BibitemOpen
  \bibfield  {author} {\bibinfo {author} {\bibfnamefont {H.}~\bibnamefont
  {Haug}}\ and\ \bibinfo {author} {\bibfnamefont {S.}~\bibnamefont {Koch}},\
  }\href@noop {} {\emph {\bibinfo {title} {{Quantum Theory of the Optical and
  Electronic Properties of Semiconductors}}}},\ \bibinfo {edition} {4th}\ ed.\
  (\bibinfo  {publisher} {World Scientific},\ \bibinfo {address} {Singapore},\
  \bibinfo {year} {2004})\ pp.\ \bibinfo {pages} {74--87}\BibitemShut {NoStop}%
\bibitem [{\citenamefont {Katsch}\ \emph {et~al.}(2020)\citenamefont {Katsch},
  \citenamefont {Selig},\ and\ \citenamefont {Knorr}}]{Katsch2020}%
  \BibitemOpen
  \bibfield  {author} {\bibinfo {author} {\bibfnamefont {F.}~\bibnamefont
  {Katsch}}, \bibinfo {author} {\bibfnamefont {M.}~\bibnamefont {Selig}}, \
  and\ \bibinfo {author} {\bibfnamefont {A.}~\bibnamefont {Knorr}},\
  }\href@noop {} {\bibfield  {journal} {\bibinfo  {journal} {2D Materials}\
  }\textbf {\bibinfo {volume} {7}},\ \bibinfo {pages} {015021} (\bibinfo {year}
  {2020})}\BibitemShut {NoStop}%
\bibitem [{\citenamefont {Hannes}\ \emph {et~al.}(2020)\citenamefont {Hannes},
  \citenamefont {Trautmann}, \citenamefont {Stein}, \citenamefont
  {Sch{\"{a}}fer}, \citenamefont {Koch},\ and\ \citenamefont
  {Meier}}]{Hannes2020}%
  \BibitemOpen
  \bibfield  {author} {\bibinfo {author} {\bibfnamefont {W.}~\bibnamefont
  {Hannes}}, \bibinfo {author} {\bibfnamefont {A.}~\bibnamefont {Trautmann}},
  \bibinfo {author} {\bibfnamefont {M.}~\bibnamefont {Stein}}, \bibinfo
  {author} {\bibfnamefont {F.}~\bibnamefont {Sch{\"{a}}fer}}, \bibinfo {author}
  {\bibfnamefont {M.}~\bibnamefont {Koch}}, \ and\ \bibinfo {author}
  {\bibfnamefont {T.}~\bibnamefont {Meier}},\ }\href {\doibase
  10.1103/PhysRevB.101.075203} {\bibfield  {journal} {\bibinfo  {journal}
  {Physical Review B}\ }\textbf {\bibinfo {volume} {101}},\ \bibinfo {pages}
  {75203} (\bibinfo {year} {2020})}\BibitemShut {NoStop}%
\bibitem [{\citenamefont {Meckbach}\ \emph {et~al.}(2018)\citenamefont
  {Meckbach}, \citenamefont {Stroucken},\ and\ \citenamefont
  {Koch}}]{Meckbach2018}%
  \BibitemOpen
  \bibfield  {author} {\bibinfo {author} {\bibfnamefont {L.}~\bibnamefont
  {Meckbach}}, \bibinfo {author} {\bibfnamefont {T.}~\bibnamefont {Stroucken}},
  \ and\ \bibinfo {author} {\bibfnamefont {S.~W.}\ \bibnamefont {Koch}},\
  }\href {\doibase 10.1103/PhysRevB.97.035425} {\bibfield  {journal} {\bibinfo
  {journal} {Physical Review B}\ }\textbf {\bibinfo {volume} {97}},\ \bibinfo
  {pages} {1} (\bibinfo {year} {2018})}\BibitemShut {NoStop}%
\bibitem [{\citenamefont {Ran}\ \emph {et~al.}(2018)\citenamefont {Ran},
  \citenamefont {Wang}, \citenamefont {Li}, \citenamefont {Yang}, \citenamefont
  {Zhao}, \citenamefont {Biswas}, \citenamefont {Singh},\ and\ \citenamefont
  {Zhang}}]{Ran2018}%
  \BibitemOpen
  \bibfield  {author} {\bibinfo {author} {\bibfnamefont {Z.}~\bibnamefont
  {Ran}}, \bibinfo {author} {\bibfnamefont {X.}~\bibnamefont {Wang}}, \bibinfo
  {author} {\bibfnamefont {Y.}~\bibnamefont {Li}}, \bibinfo {author}
  {\bibfnamefont {D.}~\bibnamefont {Yang}}, \bibinfo {author} {\bibfnamefont
  {X.-G.}\ \bibnamefont {Zhao}}, \bibinfo {author} {\bibfnamefont
  {K.}~\bibnamefont {Biswas}}, \bibinfo {author} {\bibfnamefont {D.~J.}\
  \bibnamefont {Singh}}, \ and\ \bibinfo {author} {\bibfnamefont
  {L.}~\bibnamefont {Zhang}},\ }\href {\doibase 10.1038/s41524-018-0071-1}
  {\bibfield  {journal} {\bibinfo  {journal} {npj Computational Materials}\
  }\textbf {\bibinfo {volume} {4}},\ \bibinfo {pages} {14} (\bibinfo {year}
  {2018})}\BibitemShut {NoStop}%
\bibitem [{\citenamefont {Kunioku}\ \emph {et~al.}(2016)\citenamefont
  {Kunioku}, \citenamefont {Higashi},\ and\ \citenamefont {Abe}}]{Kunioku2016}%
  \BibitemOpen
  \bibfield  {author} {\bibinfo {author} {\bibfnamefont {H.}~\bibnamefont
  {Kunioku}}, \bibinfo {author} {\bibfnamefont {M.}~\bibnamefont {Higashi}}, \
  and\ \bibinfo {author} {\bibfnamefont {R.}~\bibnamefont {Abe}},\ }\href
  {\doibase 10.1038/srep32664} {\bibfield  {journal} {\bibinfo  {journal}
  {Scientific Reports}\ }\textbf {\bibinfo {volume} {6}},\ \bibinfo {pages}
  {32664} (\bibinfo {year} {2016})}\BibitemShut {NoStop}%
\bibitem [{\citenamefont {Rossi}\ \emph {et~al.}(1999)\citenamefont {Rossi},
  \citenamefont {Goldoni}, \citenamefont {Mauritz},\ and\ \citenamefont
  {Molinari}}]{Rossi1999}%
  \BibitemOpen
  \bibfield  {author} {\bibinfo {author} {\bibfnamefont {F.}~\bibnamefont
  {Rossi}}, \bibinfo {author} {\bibfnamefont {G.}~\bibnamefont {Goldoni}},
  \bibinfo {author} {\bibfnamefont {O.}~\bibnamefont {Mauritz}}, \ and\
  \bibinfo {author} {\bibfnamefont {E.}~\bibnamefont {Molinari}},\ }\href
  {\doibase 10.1088/0953-8984/11/31/306} {\bibfield  {journal} {\bibinfo
  {journal} {Journal of Physics Condensed Matter}\ }\textbf {\bibinfo {volume}
  {11}},\ \bibinfo {pages} {5969} (\bibinfo {year} {1999})}\BibitemShut
  {NoStop}%
\bibitem [{\citenamefont {Chen}\ \emph {et~al.}(2005)\citenamefont {Chen},
  \citenamefont {Perebeinos}, \citenamefont {Freitag}, \citenamefont {Tsang},
  \citenamefont {Fu}, \citenamefont {Liu},\ and\ \citenamefont
  {Avouris}}]{Chen2005}%
  \BibitemOpen
  \bibfield  {author} {\bibinfo {author} {\bibfnamefont {J.}~\bibnamefont
  {Chen}}, \bibinfo {author} {\bibfnamefont {V.}~\bibnamefont {Perebeinos}},
  \bibinfo {author} {\bibfnamefont {M.}~\bibnamefont {Freitag}}, \bibinfo
  {author} {\bibfnamefont {J.}~\bibnamefont {Tsang}}, \bibinfo {author}
  {\bibfnamefont {Q.}~\bibnamefont {Fu}}, \bibinfo {author} {\bibfnamefont
  {J.}~\bibnamefont {Liu}}, \ and\ \bibinfo {author} {\bibfnamefont
  {P.}~\bibnamefont {Avouris}},\ }\href
  {papers://dc5b7a66-c191-4125-90d3-2536c98d4438/Paper/p453} {\bibfield
  {journal} {\bibinfo  {journal} {Science}\ }\textbf {\bibinfo {volume}
  {310}},\ \bibinfo {pages} {1171} (\bibinfo {year} {2005})}\BibitemShut
  {NoStop}%
\bibitem [{\citenamefont {Brus}(2010)}]{Brus2010}%
  \BibitemOpen
  \bibfield  {author} {\bibinfo {author} {\bibfnamefont {L.}~\bibnamefont
  {Brus}},\ }\href {\doibase 10.1021/nl904263b} {\bibfield  {journal} {\bibinfo
   {journal} {Nano Letters}\ }\textbf {\bibinfo {volume} {10}},\ \bibinfo
  {pages} {363} (\bibinfo {year} {2010})}\BibitemShut {NoStop}%
\bibitem [{\citenamefont {Lee}\ \emph {et~al.}(1995)\citenamefont {Lee},
  \citenamefont {Song}, \citenamefont {Park}, \citenamefont {Choe},
  \citenamefont {Jin},\ and\ \citenamefont {Kim}}]{Jin1995}%
  \BibitemOpen
  \bibfield  {author} {\bibinfo {author} {\bibfnamefont {Y.~L.}\ \bibnamefont
  {Lee}}, \bibinfo {author} {\bibfnamefont {H.~J.}\ \bibnamefont {Song}},
  \bibinfo {author} {\bibfnamefont {S.~A.}\ \bibnamefont {Park}}, \bibinfo
  {author} {\bibfnamefont {S.~H.}\ \bibnamefont {Choe}}, \bibinfo {author}
  {\bibfnamefont {M.~S.}\ \bibnamefont {Jin}}, \ and\ \bibinfo {author}
  {\bibfnamefont {W.~T.}\ \bibnamefont {Kim}},\ }\href {\doibase
  10.1080/00150199508228305} {\bibfield  {journal} {\bibinfo  {journal}
  {Semicond. Sci. Technol.}\ }\textbf {\bibinfo {volume} {10}},\ \bibinfo
  {pages} {1167} (\bibinfo {year} {1995})}\BibitemShut {NoStop}%
\bibitem [{\citenamefont {Demchenko}\ \emph {et~al.}(2016)\citenamefont
  {Demchenko}, \citenamefont {Izyumskaya}, \citenamefont {Feneberg},
  \citenamefont {Avrutin}, \citenamefont {{\"{O}}zg{\"{u}}r}, \citenamefont
  {Goldhahn},\ and\ \citenamefont {Morko{\c{c}}}}]{Demchenko2016}%
  \BibitemOpen
  \bibfield  {author} {\bibinfo {author} {\bibfnamefont {D.~O.}\ \bibnamefont
  {Demchenko}}, \bibinfo {author} {\bibfnamefont {N.}~\bibnamefont
  {Izyumskaya}}, \bibinfo {author} {\bibfnamefont {M.}~\bibnamefont
  {Feneberg}}, \bibinfo {author} {\bibfnamefont {V.}~\bibnamefont {Avrutin}},
  \bibinfo {author} {\bibnamefont {{\"{O}}zg{\"{u}}r}}, \bibinfo {author}
  {\bibfnamefont {R.}~\bibnamefont {Goldhahn}}, \ and\ \bibinfo {author}
  {\bibfnamefont {H.}~\bibnamefont {Morko{\c{c}}}},\ }\href {\doibase
  10.1103/PhysRevB.94.075206} {\bibfield  {journal} {\bibinfo  {journal}
  {Physical Review B}\ }\textbf {\bibinfo {volume} {94}},\ \bibinfo {pages} {1}
  (\bibinfo {year} {2016})}\BibitemShut {NoStop}%
\bibitem [{\citenamefont {Mosconi}\ \emph {et~al.}(2016)\citenamefont
  {Mosconi}, \citenamefont {Umari},\ and\ \citenamefont {{De
  Angelis}}}]{Mosconi2016}%
  \BibitemOpen
  \bibfield  {author} {\bibinfo {author} {\bibfnamefont {E.}~\bibnamefont
  {Mosconi}}, \bibinfo {author} {\bibfnamefont {P.}~\bibnamefont {Umari}}, \
  and\ \bibinfo {author} {\bibfnamefont {F.}~\bibnamefont {{De Angelis}}},\
  }\href {\doibase 10.1039/c6cp03969c} {\bibfield  {journal} {\bibinfo
  {journal} {Physical Chemistry Chemical Physics}\ }\textbf {\bibinfo {volume}
  {18}},\ \bibinfo {pages} {27158} (\bibinfo {year} {2016})}\BibitemShut
  {NoStop}%
\bibitem [{\citenamefont {Booth}\ \emph {et~al.}()\citenamefont {Booth},
  \citenamefont {Klymenko}, \citenamefont {Cole},\ and\ \citenamefont
  {Russo}}]{Booth_2021_SI}%
  \BibitemOpen
  \bibfield  {author} {\bibinfo {author} {\bibfnamefont {J.~M.}\ \bibnamefont
  {Booth}}, \bibinfo {author} {\bibfnamefont {M.~V.}\ \bibnamefont {Klymenko}},
  \bibinfo {author} {\bibfnamefont {J.~H.}\ \bibnamefont {Cole}}, \ and\
  \bibinfo {author} {\bibfnamefont {S.~P.}\ \bibnamefont {Russo}},\ }\href@noop
  {} {\bibinfo  {journal} {See Supplemental Material at [URL will be inserted
  by publisher] for a more detailed description of the SBE, BSE and TDDFT
  techniques, along with convergence tests, band structures and Tauc plots}\
  }\BibitemShut {NoStop}%
\bibitem [{\citenamefont {Mukamel}(1995)}]{mukamel1995}%
  \BibitemOpen
\bibfield  {journal} {  }\bibfield  {author} {\bibinfo {author} {\bibfnamefont
  {S.}~\bibnamefont {Mukamel}},\ }\href
  {https://books.google.com.au/books?id=k\_7uAAAAMAAJ} {\emph {\bibinfo {title}
  {Principles of Nonlinear Optical Spectroscopy}}},\ Optical and Imaging
  Sciences Series\ (\bibinfo  {publisher} {Oxford University Press},\ \bibinfo
  {year} {1995})\ p.\ \bibinfo {pages} {543}\BibitemShut {NoStop}%
\bibitem [{\citenamefont {Axt}\ and\ \citenamefont {Mukamel}(1998)}]{Axt}%
  \BibitemOpen
  \bibfield  {author} {\bibinfo {author} {\bibfnamefont {V.~M.}\ \bibnamefont
  {Axt}}\ and\ \bibinfo {author} {\bibfnamefont {S.}~\bibnamefont {Mukamel}},\
  }\href {\doibase 10.1103/RevModPhys.70.145} {\bibfield  {journal} {\bibinfo
  {journal} {Rev. Mod. Phys.}\ }\textbf {\bibinfo {volume} {70}},\ \bibinfo
  {pages} {145} (\bibinfo {year} {1998})}\BibitemShut {NoStop}%
\bibitem [{\citenamefont {Takahashi}\ and\ \citenamefont
  {Mukamel}(1994)}]{Takahashi}%
  \BibitemOpen
  \bibfield  {author} {\bibinfo {author} {\bibfnamefont {A.}~\bibnamefont
  {Takahashi}}\ and\ \bibinfo {author} {\bibfnamefont {S.}~\bibnamefont
  {Mukamel}},\ }\href@noop {} {\bibfield  {journal} {\bibinfo  {journal} {The
  Journal of Chemical Physics}\ }\textbf {\bibinfo {volume} {100}},\ \bibinfo
  {pages} {2366} (\bibinfo {year} {1994})}\BibitemShut {NoStop}%
\bibitem [{\citenamefont {Yokojima}\ \emph {et~al.}(1997)\citenamefont
  {Yokojima}, \citenamefont {Meier},\ and\ \citenamefont {Mukamel}}]{Yokojima}%
  \BibitemOpen
  \bibfield  {author} {\bibinfo {author} {\bibfnamefont {S.}~\bibnamefont
  {Yokojima}}, \bibinfo {author} {\bibfnamefont {T.}~\bibnamefont {Meier}}, \
  and\ \bibinfo {author} {\bibfnamefont {S.}~\bibnamefont {Mukamel}},\
  }\href@noop {} {\bibfield  {journal} {\bibinfo  {journal} {The Journal of
  Chemical Physics}\ }\textbf {\bibinfo {volume} {106}},\ \bibinfo {pages}
  {3837} (\bibinfo {year} {1997})}\BibitemShut {NoStop}%
\bibitem [{\citenamefont {Kira}\ and\ \citenamefont {Koch}(2011)}]{kira}%
  \BibitemOpen
  \bibfield  {author} {\bibinfo {author} {\bibfnamefont {M.}~\bibnamefont
  {Kira}}\ and\ \bibinfo {author} {\bibfnamefont {S.}~\bibnamefont {Koch}},\
  }\href {https://books.google.com.au/books?id=Io-Xlb29L7AC} {\emph {\bibinfo
  {title} {Semiconductor Quantum Optics}}}\ (\bibinfo  {publisher} {Cambridge
  University Press},\ \bibinfo {year} {2011})\ p.\ \bibinfo {pages}
  {643}\BibitemShut {NoStop}%
\bibitem [{\citenamefont {Haug}\ and\ \citenamefont
  {Koch}(2004{\natexlab{b}})}]{Haug}%
  \BibitemOpen
  \bibfield  {author} {\bibinfo {author} {\bibfnamefont {H.}~\bibnamefont
  {Haug}}\ and\ \bibinfo {author} {\bibfnamefont {S.}~\bibnamefont {Koch}},\
  }\href {https://books.google.com.au/books?id=-UoG0Hx0w04C} {\emph {\bibinfo
  {title} {Quantum Theory of the Optical and Electronic Properties of
  Semiconductors}}}\ (\bibinfo  {publisher} {World Scientific},\ \bibinfo
  {year} {2004})\ p.~\bibinfo {pages} {71}\BibitemShut {NoStop}%
\bibitem [{\citenamefont {Helgaker}\ \emph {et~al.}(2014)\citenamefont
  {Helgaker}, \citenamefont {Jorgensen},\ and\ \citenamefont {Olsen}}]{Olsen}%
  \BibitemOpen
  \bibfield  {author} {\bibinfo {author} {\bibfnamefont {T.}~\bibnamefont
  {Helgaker}}, \bibinfo {author} {\bibfnamefont {P.}~\bibnamefont {Jorgensen}},
  \ and\ \bibinfo {author} {\bibfnamefont {J.}~\bibnamefont {Olsen}},\ }\href
  {https://books.google.com.au/books?id=lNVLBAAAQBAJ} {\emph {\bibinfo {title}
  {Molecular Electronic-Structure Theory}}}\ (\bibinfo  {publisher} {Wiley},\
  \bibinfo {year} {2014})\ p.\ \bibinfo {pages} {944}\BibitemShut {NoStop}%
\bibitem [{\citenamefont {Yang}\ \emph
  {et~al.}(2015{\natexlab{b}})\citenamefont {Yang}, \citenamefont {Sottile},\
  and\ \citenamefont {Ullrich}}]{Ulrich2015}%
  \BibitemOpen
  \bibfield  {author} {\bibinfo {author} {\bibfnamefont {Z.-h.}\ \bibnamefont
  {Yang}}, \bibinfo {author} {\bibfnamefont {F.}~\bibnamefont {Sottile}}, \
  and\ \bibinfo {author} {\bibfnamefont {C.~A.}\ \bibnamefont {Ullrich}},\
  }\href {\doibase 10.1103/PhysRevB.92.035202} {\bibfield  {journal} {\bibinfo
  {journal} {Phys. Rev. B}\ }\textbf {\bibinfo {volume} {92}},\ \bibinfo
  {pages} {035202} (\bibinfo {year} {2015}{\natexlab{b}})}\BibitemShut
  {NoStop}%
\bibitem [{\citenamefont {Persson}(2014)}]{BiSBr}%
  \BibitemOpen
  \bibfield  {author} {\bibinfo {author} {\bibfnamefont {K.}~\bibnamefont
  {Persson}},\ }\href {\doibase 10.17188/1199429} {\enquote {\bibinfo {title}
  {{Materials Data on BiSBr (SG:62) by Materials Project}},}\ } (\bibinfo
  {year} {2014})\BibitemShut {NoStop}%
\bibitem [{\citenamefont {Blochl}(1994)}]{Blochl1994b}%
  \BibitemOpen
  \bibfield  {author} {\bibinfo {author} {\bibfnamefont {P.~E.}\ \bibnamefont
  {Blochl}},\ }\href@noop {} {\bibfield  {journal} {\bibinfo  {journal} {Phys.
  Rev. B}\ }\textbf {\bibinfo {volume} {50}},\ \bibinfo {pages} {17953}
  (\bibinfo {year} {1994})}\BibitemShut {NoStop}%
\bibitem [{\citenamefont {Kresse}\ and\ \citenamefont
  {Furthm{\"{u}}ller}(1996)}]{Kresse1996}%
  \BibitemOpen
  \bibfield  {author} {\bibinfo {author} {\bibfnamefont {G.}~\bibnamefont
  {Kresse}}\ and\ \bibinfo {author} {\bibfnamefont {J.}~\bibnamefont
  {Furthm{\"{u}}ller}},\ }\href@noop {} {\bibfield  {journal} {\bibinfo
  {journal} {Phys. Rev. B}\ }\textbf {\bibinfo {volume} {54}},\ \bibinfo
  {pages} {11169} (\bibinfo {year} {1996})}\BibitemShut {NoStop}%
\bibitem [{\citenamefont {Perdew}\ \emph {et~al.}(1996)\citenamefont {Perdew},
  \citenamefont {Burke},\ and\ \citenamefont {Ernzerhof}}]{Perdew1996}%
  \BibitemOpen
  \bibfield  {author} {\bibinfo {author} {\bibfnamefont {J.~P.}\ \bibnamefont
  {Perdew}}, \bibinfo {author} {\bibfnamefont {K.}~\bibnamefont {Burke}}, \
  and\ \bibinfo {author} {\bibfnamefont {M.}~\bibnamefont {Ernzerhof}},\
  }\href@noop {} {\bibfield  {journal} {\bibinfo  {journal} {Phys. Rev. Lett.}\
  }\textbf {\bibinfo {volume} {77}},\ \bibinfo {pages} {3865} (\bibinfo {year}
  {1996})}\BibitemShut {NoStop}%
\bibitem [{\citenamefont {Heyd}\ \emph {et~al.}(2003)\citenamefont {Heyd},
  \citenamefont {Scuseria},\ and\ \citenamefont {Ernzerhof}}]{Heyd2003}%
  \BibitemOpen
  \bibfield  {author} {\bibinfo {author} {\bibfnamefont {J.}~\bibnamefont
  {Heyd}}, \bibinfo {author} {\bibfnamefont {G.~E.}\ \bibnamefont {Scuseria}},
  \ and\ \bibinfo {author} {\bibfnamefont {M.}~\bibnamefont {Ernzerhof}},\
  }\href {\doibase 10.1063/1.1564060} {\bibfield  {journal} {\bibinfo
  {journal} {J. Chem. Phys.}\ }\textbf {\bibinfo {volume} {118}},\ \bibinfo
  {pages} {8207} (\bibinfo {year} {2003})}\BibitemShut {NoStop}%
\bibitem [{\citenamefont {Monkhorst}\ and\ \citenamefont
  {Pack}(1976)}]{Monkhorst1976}%
  \BibitemOpen
  \bibfield  {author} {\bibinfo {author} {\bibfnamefont {H.~J.}\ \bibnamefont
  {Monkhorst}}\ and\ \bibinfo {author} {\bibfnamefont {J.~D.}\ \bibnamefont
  {Pack}},\ }\href@noop {} {\bibfield  {journal} {\bibinfo  {journal} {Phys.
  Rev. B}\ }\textbf {\bibinfo {volume} {13}},\ \bibinfo {pages} {5188}
  (\bibinfo {year} {1976})}\BibitemShut {NoStop}%
\bibitem [{\citenamefont {Shishkin}\ and\ \citenamefont
  {Kresse}(2006)}]{Shishkin2006}%
  \BibitemOpen
  \bibfield  {author} {\bibinfo {author} {\bibfnamefont {M.}~\bibnamefont
  {Shishkin}}\ and\ \bibinfo {author} {\bibfnamefont {G.}~\bibnamefont
  {Kresse}},\ }\href {\doibase 10.1103/PhysRevB.74.035101} {\bibfield
  {journal} {\bibinfo  {journal} {Phys. Rev. B}\ }\textbf {\bibinfo {volume}
  {74}},\ \bibinfo {pages} {35101} (\bibinfo {year} {2006})}\BibitemShut
  {NoStop}%
\bibitem [{\citenamefont {Blochl}\ \emph {et~al.}(1994)\citenamefont {Blochl},
  \citenamefont {Jepsen},\ and\ \citenamefont {Andersen}}]{Bloechl1994}%
  \BibitemOpen
  \bibfield  {author} {\bibinfo {author} {\bibfnamefont {P.~E.}\ \bibnamefont
  {Blochl}}, \bibinfo {author} {\bibfnamefont {O.}~\bibnamefont {Jepsen}}, \
  and\ \bibinfo {author} {\bibfnamefont {O.~K.}\ \bibnamefont {Andersen}},\
  }\href@noop {} {\bibfield  {journal} {\bibinfo  {journal} {Phys. Rev. B}\
  }\textbf {\bibinfo {volume} {49}},\ \bibinfo {pages} {16223} (\bibinfo {year}
  {1994})}\BibitemShut {NoStop}%
\bibitem [{\citenamefont {Methfessel}\ and\ \citenamefont
  {Paxton}(1989)}]{Methfessel1989}%
  \BibitemOpen
  \bibfield  {author} {\bibinfo {author} {\bibfnamefont {M.}~\bibnamefont
  {Methfessel}}\ and\ \bibinfo {author} {\bibfnamefont {A.~T.}\ \bibnamefont
  {Paxton}},\ }\href@noop {} {\bibfield  {journal} {\bibinfo  {journal}
  {Physical review. B}\ }\textbf {\bibinfo {volume} {40}},\ \bibinfo {pages}
  {3616} (\bibinfo {year} {1989})}\BibitemShut {NoStop}%
\bibitem [{\citenamefont {Tauc}\ \emph {et~al.}(1966)\citenamefont {Tauc},
  \citenamefont {Grigorovici},\ and\ \citenamefont {Vancu}}]{Tauc1966}%
  \BibitemOpen
  \bibfield  {author} {\bibinfo {author} {\bibfnamefont {J.}~\bibnamefont
  {Tauc}}, \bibinfo {author} {\bibfnamefont {R.}~\bibnamefont {Grigorovici}}, \
  and\ \bibinfo {author} {\bibfnamefont {A.}~\bibnamefont {Vancu}},\
  }\href@noop {} {\bibfield  {journal} {\bibinfo  {journal} {Phys. Status
  Solidi}\ }\textbf {\bibinfo {volume} {15}},\ \bibinfo {pages} {627} (\bibinfo
  {year} {1966})}\BibitemShut {NoStop}%
\bibitem [{\citenamefont {Erben}\ \emph {et~al.}(2018)\citenamefont {Erben},
  \citenamefont {Steinhoff}, \citenamefont {Gies}, \citenamefont
  {Sch{\"{o}}nhoff}, \citenamefont {Wehling},\ and\ \citenamefont
  {Jahnke}}]{Erben2018}%
  \BibitemOpen
  \bibfield  {author} {\bibinfo {author} {\bibfnamefont {D.}~\bibnamefont
  {Erben}}, \bibinfo {author} {\bibfnamefont {A.}~\bibnamefont {Steinhoff}},
  \bibinfo {author} {\bibfnamefont {C.}~\bibnamefont {Gies}}, \bibinfo {author}
  {\bibfnamefont {G.}~\bibnamefont {Sch{\"{o}}nhoff}}, \bibinfo {author}
  {\bibfnamefont {T.~O.}\ \bibnamefont {Wehling}}, \ and\ \bibinfo {author}
  {\bibfnamefont {F.}~\bibnamefont {Jahnke}},\ }\href@noop {} {\bibfield
  {journal} {\bibinfo  {journal} {Phys. Rev. B}\ }\textbf {\bibinfo {volume}
  {98}},\ \bibinfo {pages} {035434} (\bibinfo {year} {2018})}\BibitemShut
  {NoStop}%
\bibitem [{\citenamefont {Engel}\ and\ \citenamefont
  {Farid}(1993)}]{Engel1993}%
  \BibitemOpen
  \bibfield  {author} {\bibinfo {author} {\bibfnamefont {G.~E.}\ \bibnamefont
  {Engel}}\ and\ \bibinfo {author} {\bibfnamefont {B.}~\bibnamefont {Farid}},\
  }\href@noop {} {\bibfield  {journal} {\bibinfo  {journal} {Phys Rev B}\
  }\textbf {\bibinfo {volume} {47}},\ \bibinfo {pages} {15931} (\bibinfo {year}
  {1993})}\BibitemShut {NoStop}%
\end{thebibliography}%

\end{document}